\documentclass[11pt]{article}
\usepackage{axodraw}
\usepackage{epsfig}
\usepackage{amsfonts}
\usepackage{amsmath}
\usepackage{bm,bbm}
\usepackage{cite}
\allowdisplaybreaks[1]

\input pix.sty

\newcommand{\Hfast}{{\widehat H}_\rmi{fast}}

\newcommand{\lnf}{l^{ }_\rmi{1f}}
\newcommand{\lif}{l^{ }_\rmi{2f}}
\newcommand{\lnb}{l^{ }_\rmi{1b}}
\newcommand{\lib}{l^{ }_\rmi{2b}}

\newcommand{\I}{\rmii{$I$}}
\newcommand{\J}{\rmii{$J$}}
\newcommand{\sL}{\rmii{$L$}}

\newcommand{\tmuBL}{\tilde{\mu}_\rmii{$B$+$L$}}

\newcommand{\YBmL}{Y_\rmii{$B$-$L$}}
\newcommand{\YLmB}{Y_\rmii{$L$-$B$}}
\newcommand{\YBpL}{Y_\rmii{$B$+$L$}}
\newcommand{\YB}{Y_\rmii{$B$}}
\newcommand{\YL}{Y_\rmii{$L$}}
\newcommand{\muY}{\mu_\rmii{$Y$}}
\newcommand{\bmuY}{\bar{\mu}_\rmii{$Y$}}

\newcommand{\muH}{\mu_\rmii{$H$}}
\newcommand{\muLa}{\mu_{\rmii{$L$}a}}

\newcommand{\nG}{n_\rmii{$G$}}

\newcommand{\mZ}{m_\rmii{$Z$}}

\newcommand{\kT}{k_\rmii{$T$}}
\newcommand{\km}{k_-}
\newcommand{\kp}{k_+}

\newcommand{\aL}{a^{ }_\rmii{L}}
\newcommand{\aR}{a^{ }_\rmii{R}}
\renewcommand{\eq}{eq.~}
\renewcommand{\eqs}{eqs.~}
\renewcommand{\se}{sec.~}
\renewcommand{\ses}{secs.~}
\renewcommand{\fig}{fig.~}

\newcommand{\tinymsbar}{{\overline{\mbox{\tiny\rm{MS}}}}}
\newcommand{\Lambdamsbar}{{\Lambda_\tinymsbar}}

\newcommand{\alphas}{\alpha_{\rm s}}
\newcommand{\alphaw}{\alpha_{\rm w}}

\newcommand{\Tc}{T_{\rm c}}

\newcommand{\gammaE}{\gamma_\rmii{E}}

\newcommand{\rmO}{{\mathcal{O}}}
\newcommand{\bmu}{\bar\mu}

\def\lsi{\raise0.3ex\hbox{$<$\kern-0.75em\raise-1.1ex\hbox{$\sim$}}}
\def\gsi{\raise0.3ex\hbox{$>$\kern-0.75em\raise-1.1ex\hbox{$\sim$}}}
\newcommand{\lsim}{\mathop{\lsi}}
\newcommand{\gsim}{\mathop{\gsi}}

\newcommand{\nF}{n_\rmii{F}}
\newcommand{\nB}{n_\rmii{B}}

\newcommand{\rmii}[1]{{\mbox{\tiny\rm{#1}}}}

\newcommand{\re}{\mathop{\mbox{Re}}}
\newcommand{\im}{\mathop{\mbox{Im}}}

\newcommand{\Tint}[1]{{\hbox{$\sum$}\!\!\!\!\!\!\!\int\,}_{\!\!\!\!\raise-0.9ex\hbox{$\scriptstyle{#1}$}}}
\newcommand{\Tinti}[1]{{{\Sigma}\!\!\!\!\raise0.3ex\hbox{$\int$}_\rmii{${#1}$}}}

\newcommand{\unit}{{\mathbbm{1}}} 
\newcommand{\bi}{\begin{itemize}}
\newcommand{\ei}{\end{itemize}}
\newcommand{\hide}[1]{ }
\newcommand{\bsl}[1]{\,\slash\!\!\!\!{#1}\,}
\newcommand{\msl}[1]{\,\slash\!\!\!{#1}\,}
\def\TAsc(#1,#2)(#3,#4,#5)%
{\SetWidth{2.0}\CArc(#1,#2)(#3,#4,#5)\SetWidth{1.0}}
\def\Lwidth{3}

\def\TAgl(#1,#2)(#3,#4,#5){\SetWidth{2.0}\PhotonArc(#1,#2)(#3,#4,#5){\Lwidth}%
{6.283 #3 mul 360 div #4 #5 sub #4 #5 sub mul sqrt mul Tdensity mul}%
\SetWidth{1.0}}
\def\TLgl(#1,#2)(#3,#4){\SetWidth{2.0}\Photon(#1,#2)(#3,#4){\Lwidth}
{#1 #3 sub #1 #3 sub mul #2 #4 sub #2 #4 sub mul add sqrt Tdensity mul}%
\SetWidth{1.0}}

\renewcommand{\picc}[1]{\;\parbox[c]{60pt}{\begin{picture}(60,30)(0,0)
\SetWidth{1.0}\SetScale{1.3} #1 \end{picture}}\; }
\def\Lwidth{1.3}

\renewcommand{\pic}[1]{\;\parbox[c]{30pt}{\begin{picture}(30,30)(0,-3)
\SetWidth{1.0}\SetScale{0.8} #1 \end{picture}}\;}
\renewcommand{\picb}[1]{\;\parbox[c]{45pt}{\begin{picture}(45,30)(0,-3)
\SetWidth{1.0}\SetScale{0.8} #1 \end{picture}}\;}

\makeatletter \@addtoreset{equation}{section} \makeatother
\renewcommand{\theequation}{\arabic{section}.\arabic{equation}}
\makeatletter
\renewcommand\section{\@startsection {section}{1}{\z@}%
                                   {-5.5ex \@plus -1ex \@minus -.2ex}
                                   {2.3ex \@plus.2ex}%
                                   {\normalfont\large\bfseries}}
\renewcommand\subsection{\@startsection{subsection}{2}{\z@}%
                                     {-3.25ex\@plus -1ex \@minus -.2ex}%
                                     {1.5ex \@plus .2ex}%
                                     {\normalfont\normalsize\bfseries}}
\renewcommand\thesection {\@arabic\c@section}
\renewcommand\thesubsection   {\thesection.\@arabic\c@subsection}
\renewcommand{\@seccntformat}[1]{%
\csname the#1\endcsname.\hspace{1.0em}}
\makeatother

\begin{document}

\flushbottom

\begin{titlepage}

\begin{flushright}
CERN-TH-2018-022 \\
February 2018
\vspace*{0.5cm}
\end{flushright}
\begin{centering}
\vfill

{\Large{\bf
GeV-scale hot sterile neutrino oscillations: \\[3mm] 
a numerical solution
}} 

\vspace{0.8cm}

J.~Ghiglieri$^\rmi{a}$ and M.~Laine$^\rmi{b}$

\vspace{0.8cm}

$^\rmi{a}${\em
Theoretical Physics Department, CERN, \\ 
CH-1211 Geneva 23, Switzerland \\}

\vspace{0.3cm}

$^\rmi{b}${\em
AEC, 
Institute for Theoretical Physics, 
University of Bern, \\ 
Sidlerstrasse 5, CH-3012 Bern, Switzerland \\}

\vspace*{0.8cm}

\mbox{\bf Abstract}

\end{centering}

\vspace*{0.3cm}

\noindent
The scenario of baryogenesis through GeV-scale sterile neutrino oscillations
is governed by non-linear differential equations for the time evolution 
of a sterile neutrino density matrix and Standard Model lepton and baryon
asymmetries. By employing up-to-date rate coefficients and 
a non-perturbatively estimated Chern-Simons diffusion rate, we present 
a numerical \mbox{solution} of this system,  
incorporating the full momentum and helicity dependences 
of the density matrix.  The density matrix deviates significantly 
from kinetic equilibrium, with the IR modes equilibrating much faster than 
the UV modes. For equivalent input parameters, our final results differ
moderately ($\sim 50\%$) from recent benchmarks in the literature.
The possibility of producing an observable baryon 
asymmetry is nevertheless confirmed. We illustrate the dependence of the 
baryon asymmetry on the sterile neutrino mass splitting and on the 
CP-violating phase measurable in active neutrino oscillation experiments.

\vfill

 
\vspace*{1cm}
 
\vfill

\end{titlepage}

%
\section{Introduction}
\la{se:intro}

Explaining the matter-antimatter asymmetry of the Universe through
experimentally verifiable laws of nature remains one of the most important
open issues for particle physics and cosmology. 
The scenario of baryogenesis through GeV-scale sterile neutrino oscillations
has established itself as a nice framework in which concrete progress
can be made on all aspects of this problem. 
The original idea was put forward in ref.~\cite{ars}, 
and a significant reformulation, constituting the current 
understanding of various 
parametric dependences, was provided by ref.~\cite{as}. 
Representative examples of recent refinements can be found 
in refs.~\cite{new_x,shintaro,dg,canetti,shuve,abada,val,ht1,n2,n3,es,asaka,
ht,new_y,shintaro_new,new2}. Among these, the present investigation 
can most easily be contrasted with ref.~\cite{n3}, 
whose benchmark point we adopt as a central test case 
for our numerical solution. 

The present paper is 
a follow-up to ref.~\cite{cptheory}, in which
rates and rate equations were 
derived for the behaviour of baryon and 
lepton asymmetries and the sterile neutrino density matrix
at complete leading order in Standard Model couplings.
The derivation generalized and built up on   
techniques developed in several 
previous works~\cite{bb1,bb2,kubo,interpolation,dmpheno,bsw,broken}. 
In particular it required a resummation of infrared sensitive 
$1+n \leftrightarrow 2+n$ scatterings as well as a computation
of all $2\leftrightarrow 2$ contributions to sterile
neutrino production rates 
and chemical and kinetic equilibration coefficients. 
These coefficients display a non-trivial momentum 
dependence, which in combination with the general structure
of the rate equations leads to non-trivial momentum dependences
of different components of the density matrix as well. 

The parameter space of the (type-I seesaw) model in question
has been nicely delineated in ref.~\cite{canetti}.
In a so-called ``scenario~I'', 
two sterile neutrinos are responsible for generating active
neutrino mass differences, the observed baryon asymmetry, and a large lepton 
asymmetry. A third sterile neutrino constitutes keV scale dark matter, 
whose production is resonantly boosted by the
above-mentioned large lepton asymmetry. 
In a broader ``scenario~II'', the production of a large lepton asymmetry 
is not considered, 
but the focus is otherwise on the same two-flavour problem for 
active neutrino mass differences and baryon asymmetry. 
In the parametrically most relaxed 
``scenario~III'', three flavours of sterile neutrinos 
participate in the production of active neutrino
mass differences and the baryon asymmetry. 
In a technical sense, our study corresponds to scenario~II, 
which is minimal in the {\em dimension} of its parameter space. 
However, the 
same methods would also permit to address the more restrictive 
scenario~I if the solutions for the lepton asymmetries were followed deep 
into the Higgs phase, and the more relaxed scenario~III if 
a larger-dimensional density matrix were considered. We postpone these 
numerically more demanding investigations into future.  

The structure of this paper is as follows. 
The basic equations from ref.~\cite{cptheory}, transcribed
into an expanding cosmological background, are reviewed in \ref{se:basic}. 
The most important terms, helpful for analytic understanding and 
numerical estimates, are identified in \se\ref{se:power}. The main numerical
challenge of the problem, namely that both ``fast'' and ``slow'' processes
play a role,  
is tackled in \se\ref{se:fast}. Numerical solutions are presented
in \se\ref{se:numerics}, and we conclude in \se\ref{se:concl}. 
Appendix A reviews the definitions and some relevant 
properties of the rate coefficients $Q,R,S$ from ref.~\cite{cptheory}, 
appendix B explains our parametrization
of neutrino Yukawa couplings, and  
appendix C summarizes our treatment of the so-called sphaleron rate. 

%
\section{Review of basic equations}
\la{se:basic}

We start by rewriting and completing 
the set of rate equations derived in ref.~\cite{cptheory}, 
transcribing them from a flat to an expanding background. The expansion is 
characterized by a Hubble rate
$
 H 
 = \sqrt{{8\pi e}} /(\sqrt{3} m_\rmi{Pl}^{ })
$, 
where $e$ is the energy density and 
$m^{ }_\rmi{Pl} = 1.22 \times 10^{19}$~GeV
is the Planck mass. 
The entropy density $s$
and the speed of sound squared 
$c_s^2 = \partial p / \partial e$ also appear, 
where $p$ is the pressure. Yield parameters are defined as 
\be
  Y^{ }_i \;\equiv\; \frac{n^{ }_i}{s}
 \;, 
\ee
where the $n^{ }_i$ stand for various particle number asymmetries
(``particles minus antiparticles''). 
The coefficient functions $A,B,...$ introduced 
in ref.~\cite{cptheory} are rescaled as 
\be
 \widehat A \; \equiv \; \frac{A}{3 c_s^2 H}
 \;, \quad
 \mbox{etc}
 \;. \la{widehat}
\ee
Denoting furthermore
\be
 Y' \equiv \frac{ {\rm d}Y }{ {\rm d}x } 
 \;, \quad
 x \; \equiv \; \ln \biggl( \frac{T^{ }_\rmi{max}}{T}\biggr)
 \;, \quad
 \kT^{ } \; \equiv \; 
 k \, \biggl\{ \frac{s(T)}{s(T^{ }_\rmi{min})} \biggr\}^{1/3} 
 \;, \la{def_x}
\ee
where 
$T^{ }_\rmi{max}$ is a maximal temperature, 
$T^{ }_\rmi{min}$ is a minimal temperature, 
$\kT^{ }$
is a co-moving momentum, and $k$ is the momentum at $ T = T^{ }_\rmi{min}$, 
the evolution equation for
lepton asymmetry of generation $a$ minus one third of baryon asymmetry reads
\be
 {Y}_a' - \frac{\YB'}{3} 
 \; = \;  
 \frac{4}{s} \int_{\vec{k}^{ }_\rmii{$T$}} 
 \tr \Bigl\{ 
  - 
  \, \nF(\kT^{ }) [1 - \nF(\kT^{ })] \,  \widehat A^{+}_{(a)}
   +   
   \bigl[\, \rho^{+}_{ } - \unit \, \nF(\kT^{ }) \, \bigr]
  \widehat B^{+}_{(a)} + \rho^{-}_{ } \widehat B^{-}_{(a)}
 \Bigr\} 
 \;, 
 \la{summary_na}  
\ee
where 
\ba
 \widehat A^{+}_{(a)\I\J}
 & \equiv & 
  \re (h^{ }_{\I a}h^*_{\J a}) \, 
 \bar{\mu}^{ }_a \,
 \widehat Q^{+}_{\{\I\J\}} 
 \;, \la{A} \\ 
 \widehat B^{+}_{(a)\I\J}
 & \equiv & 
 - i \im (h^{ }_{\I a}h^*_{\J a}) \, 
    \widehat Q^{+}_{\{\I\J\}} 
  + 
  \re (h^{ }_{\I a}h^*_{\J a}) \, 
  \Bigl[ 
  \bar{\mu}^{ }_a \, 
        \widehat R^{+}_{\{ \I \J\} } 
   + \bmuY^{ } \,
        \widehat S^{+}_{\{ \I\J\} } 
  \Bigr]
 \;, \la{Bplus} \\
 \widehat B^{-}_{(a)\I\J}
  & \equiv &
  \re (h^{ }_{\I a}h^*_{\J a}) \, 
    \widehat Q^{-}_{\{\I\J\}} 
 - i \im (h^{ }_{\I a}h^*_{\J a}) \, 
  \Bigl[ 
    \bar{\mu}^{ }_a \, 
     \widehat R^{-}_{\{ \I \J\} } 
   + \bmuY^{ } \, 
        \widehat S^{-}_{\{ \I\J\} }
  \Bigr]
 \;. \la{Bminus}
\ea
Here $\widehat Q$, $\widehat R$ and $\widehat S$ are rate 
coefficients from ref.~\cite{cptheory} that 
have been rescaled as in \eq\nr{widehat};\footnote{%
 The basic definitions of 
 $Q,R$ and $S$, some of their relevant properties, 
 and an update on their numerical evaluation,
 are summarized in appendix~A. 
 }  
$\rho^{\pm}_{ }$ are helicity-symmetrized and antisymmetrized
density matrices; $h^{ }_{\I a} \equiv (h^{ }_{\nu})^{ }_{\I a}$ 
are Yukawas coupling a sterile
neutrino of flavour $I$ to an active lepton of generation $a$; 
$\bar\mu^{ }_a \equiv \mu^{ }_a/T$  
and $\bmuY \equiv \muY/T$ are rescaled lepton and hypercharge chemical
potentials;\footnote{%
 The latter represents, more properly, 
 the expectation value of the hypercharge gauge potential. 
 } 
and unexplained notation is identical to that
in ref.~\cite{cptheory}.  
A way to fix the values of $h^{ }_{\I a}$ in terms of observable
quantities is reviewed in appendix~B.

The evolution equations of the density matrices, integrated along 
co-moving momenta, read
\ba
  ({\rho}^{\pm}_{ })'(\kT^{ })  & = & 
   i \bigl[\widehat H^{ }_{0}, \rho^{\pm}_{ } \bigr] 
   + 
   i \bigl[\widehat \Delta^{ }_{0}, \rho^{\mp}_{ } \bigr] 
   + 2 \nF(\kT^{ }) [ 1-  \nF(\kT^{ })] \, \widehat C^{\pm}_{ } \nn[2mm] 
   & - & 
   \widehat D^{\pm}_{ }  
   \bigl[ \rho^{+}_{ } - \unit \nF(\kT^{ }) \bigr]
  - 
   \bigl[ \rho^{+}_{ } - \unit \nF(\kT^{ }) \bigr]  
   \widehat D^{\pm\dagger}_{ }
  - \widehat D^{\mp}_{ }  \rho^{-}_{ }
  - \rho^{-}_{ }  \widehat D^{\mp\dagger}_{ }
 \;. \la{summary_rho_minus}
\ea
The coefficients describing real processes
(particle creations and annihilations) are
\ba
 \widehat C^{+}_{\I\J} & \equiv & 
 - i {\textstyle\sum_a} 
   \im (h^{ }_{\I a} h^{*}_{\J a})\, \bar{\mu}^{ }_a  
   \, \widehat Q^{+}_{\{\I\J\} }
 \;, \la{Cplus} \\[2mm] 
 \widehat C^{-}_{\I\J} & \equiv & 
 {\textstyle\sum_a} 
 \re (h^{ }_{\I a} h^{*}_{\J a})\, \bar{\mu}^{ }_a  
   \, \widehat Q^{-}_{\{\I\J\} }
 \;, \la{Cminus} \\[2mm] 
 \widehat D^{+}_{\I\J} & \equiv & 
 {\textstyle\sum_a}
   \re (h^{ }_{\I a} h^{*}_{\J a})   
   \,  
   \widehat Q^{+}_{\I\J}
  -  i \, {\textstyle\sum_a} \im (h^{ }_{\I a} h^{*}_{\J a})  \, 
  \Bigl[ \bar{\mu}^{ }_a \, \widehat R^{+}_{\I\J}
   + \bmuY^{ } \widehat S^{+}_{\I\J}
  \Bigr] \,  
 \;, \la{Dplus}  \\[2mm]
 \widehat D^{-}_{\I\J} & \equiv & 
  -  i \, {\textstyle\sum_a} \im (h^{ }_{\I a} h^{*}_{\J a})  \, 
   \widehat Q^{-}_{\I\J}
  + 
 {\textstyle\sum_a}
   \re (h^{ }_{\I a} h^{*}_{\J a})   
   \,  
  \Bigl[ \bar{\mu}^{ }_a \, \widehat R^{-}_{\I\J}
   + \bmuY^{ } \widehat S^{-}_{\I\J}
  \Bigr]
 \;, \la{Dminus}  
\ea
whereas the unitary part of the evolution 
is determined by a Hermitean Hamiltonian with 
\ba
 \widehat H^{ }_{0\I\J} & = & \frac{1}{6 \kT^{ } c_s^2 H} \biggl\{
 \delta^{ }_{\I\J} 
 \biggl[ M_\I^2 - \frac{ \sum_{\sL}
  (M_\sL^2 + \sfr14 \sum_a |h^{ }_{\sL a}|^2 T^2) }{\sum_{\sL}}  \biggr]
 + 
 \frac{{\textstyle\sum_a} \re (
     h^{ }_{\I a}h^{*}_{\J a} ) T^2
   }{4}  \biggr\}  
 \;, \hspace*{7mm} \la{H0} \\[2mm]
 \widehat \Delta^{ }_{0\I\J} & = &  
 - 
 \frac{i\, {\textstyle\sum_a} \im(
     h^{ }_{\I a}h^{*}_{\J a}  ) T^2
   }{24\kT^{ } c_s^2 H }    
 \;. \la{Delta0}
\ea
We have here chosen $\widehat H^{ }_0$ to be traceless 
(the trace part drops out 
in \eq\nr{summary_rho_minus}).

A further rate equation concerns the time evolution of the baryon asymmetry, 
and requires a careful discussion. Let us denote the 
right-hand side of \eq\nr{summary_na} as a ``force'', $F^{ }_a$.
If we were to write equations separately for 
$Y^{ }_a$ and $Y^{ }_\rmii{$B$}$, they would have the forms 
\ba
 Y_a' & = & F_a^{ } + \frac{F^{ }_\rmi{diff}}{6}  
 \;, \la{Yap} \\ 
 \YB' & = & \frac{F^{ }_\rmi{diff}}{2} 
 \;,
\ea
where $F^{ }_\rmi{diff}$ is the anomalous
baryon plus lepton number violating rate. 
Going over to the usual variables $Y^{ }_a - \YB^{ }/3$ and 
$\YBpL^{ } \equiv \sum_a Y^{ }_a + \YB^{ }$, 
the rate equations become
\ba
 Y_a' - \frac{\YB'}{3} 
 & = & F_a^{ } 
 \;, \\ 
 \YBpL' & = &
 \sum_a  F_a^{ } + F^{ }_\rmi{diff}
 \;. \la{YBLp}
\ea
At high temperatures, where $ F^{ }_\rmi{diff} \gg \sum_a  F_a^{ }$, 
the first term is sometimes omitted from \eq\nr{YBLp}
(cf.\ e.g.\ ref.~\cite{cptheory}). 
However, we want to solve the equations down 
to low temperatures, where
$ F^{ }_\rmi{diff} \ll \sum_a  F_a^{ }$, and then this 
term must be kept. It guarantees that the baryon yield 
stops evolving below the electroweak crossover: 
\be
 \YB' 
 \; = \;
 \frac{\YBpL' - \sum_a [Y_a' - \YB'/3]}{2}
 \; = \; 
 \frac{F^{ }_\rmi{diff}}{2} 
 \;. 
\ee
Following the notation of ref.~\cite{cptheory}, 
the anomalous force term here reads ($\nG^{ }\equiv 3$)
\be
 F^{ }_\rmi{diff} 
 \; = \;
 - \frac {2 \nG^2\, \Gamma^{ }_\rmi{diff}(T)}
         {3 s c_s^2 H } \, \frac{\tmuBL }{ T}
 \;, \la{summary_nB}
\ee
where $\tmuBL^{ }$ is a chemical potential associated
with the baryon plus lepton asymmetry, and $\Gamma^{ }_\rmi{diff}$ 
is the Chern-Simons diffusion coefficient, whose 
$T$-dependence is reviewed in appendix~C.

The equations above depend on the chemical potentials
$\bar\mu^{ }_a$, $\bmuY$ and $\tmuBL$.
The first two can be obtained by 
going through chemical potentials associated with 
lepton minus baryon asymmetries, $\tilde\mu^{ }_a$, 
and through $\tmuBL^{ }$, via~\cite{cptheory}
\ba
  \bar \mu^{ }_a & = & \frac{ \tilde{\mu}^{ }_a + \tmuBL^{ } }{T}  
 \;, \la{bmua} \\ 
  \bmuY^{ } & = & \frac{8}{33 T} \Bigl( \sum_a \tilde{\mu}^{ }_a +
 \frac{ 3 \tmuBL^{ }}{2} \Bigr) 
 \;. 
\ea
Here, 
up to corrections of $\rmO(\alphaw^{1/2},\alphas)$~\cite{kubo,sangel}, 
\ba 
 \left( 
  \begin{array}{c}
    \tilde{\mu}^{ }_1 \\
    \tilde{\mu}^{ }_2 \\ 
    \tilde{\mu}^{ }_3 \\
    \tmuBL^{ }   
  \end{array}
 \right)
 & = & \frac{1}{144 T^2} 
 \left( 
   \begin{array}{rrrr} 
     319 & 31 & 31 & -23 \\  
     31 & 319 & 31 & -23 \\ 
     31 & 31 & 319 & -23 \\
     -23 & -23 & -23 & 79
   \end{array}
 \right)
 \left( 
  \begin{array}{c}
    n^{ }_1 - \frac{\nB^{ }}{3}  \\
    n^{ }_2 - \frac{\nB^{ }}{3} \\ 
    n^{ }_3 - \frac{\nB^{ }}{3} \\
   \nB + \sum_a n^{ }_a  
  \end{array} 
 \right) 
 \;. \la{mu_lowT}
\ea
This closes the set of rate equations. (The matrix
appearing in \eq\nr{mu_lowT} is
modified in the Higgs phase~\cite{khlebnikov}, 
but for our considerations 
at $T \gsim 130$~GeV where the Higgs expectation value 
is parametrically $v \lsim gT$, 
this amounts to a higher-order effect.) 

%
\section{Identification of the most important terms}
\la{se:power}

In order to solve 
the equations of \se\ref{se:basic} numerically, 
it is convenient to go over into the interaction picture. 
Moreover, in order to understand the structure of the solution, 
it is helpful to identify which of the many terms on the right-hand
sides of the equations are the most important ones. The latter maneuver 
is not necessary for a numerical solution at early times, however 
it facilitates finding a simplified solution valid at late times
(cf.\ \se\ref{ss:osc}).  

As a first step, 
focussing for concreteness on two generations, we rename 
the upper diagonal component of $\widehat H^{ }_0$, 
i.e.\  $(\widehat H^{ }_0)^{ }_{11}$, as
\be
 \Hfast^{ } \; \equiv \; 
 \frac{1}{12\kT^{ }c_s^2 H}
 \biggl[ 
 M_1^2 - M_2^2 + \frac{\sum_a(|h^{ }_{1a}|^2 - |h^{ }_{2a}|^2)T^2}{4}
 \biggr]
 \;. \la{Hfast}
\ee
The essential term here is the vacuum mass difference
$M_1^2 - M_2^2$.
Let $U$ be a rapidly varying phase factor satisfying 
\be
 U'(x) = i \Hfast^{ }(x)\, U(x)
 \;,  \la{U_def}
\ee
and denote 
\be
 \rho^{\pm}_{ } \; \equiv \; 
 \biggl(
 \begin{array}{cc} 
   U & 0 \\ 
   0 & U^*  
 \end{array}
 \biggr)
 \widetilde \rho\,^{\pm}_{ }
 \biggl(
 \begin{array}{cc} 
   U^* & 0 \\ 
   0 & U  
 \end{array}
 \biggr)
 \;, \quad 
 \widehat A^{+}_{(a)} \; \equiv \; 
 \biggl(
 \begin{array}{cc} 
   U & 0 \\ 
   0 & U^*  
 \end{array}
 \biggr)
 \widetilde A^{+}_{(a)}
 \biggl(
 \begin{array}{cc} 
   U^* & 0 \\ 
   0 & U  
 \end{array}
 \biggr)
 \;, \la{dirac}
\ee
and similarly for the other coefficients. 
Substituting \eq\nr{dirac} into \eqs\nr{summary_na}
and \nr{summary_rho_minus}, 
the equations of motion retain their form but with 
$\widehat A$ replaced by 
$\widetilde A$, etc, 
and $\widehat H^{ }_0$ replaced by 
$\widetilde H_\rmi{slow}^{ } \equiv 
 \widetilde H^{ }_0 - 
 \mathop{\mbox{diag}}(\Hfast^{ },
 -\Hfast^{ })$. 
Simultaneously the off-diagonal components of the 
coefficient matrices become time-dependent: 
\be
 \widetilde A = 
 \biggl(
 \begin{array}{cc} 
   U^* & 0 \\ 
   0 & U  
 \end{array}
 \biggr)
 \biggl(
 \begin{array}{cc} 
  \widehat A^{ }_{11} &  \widehat A^{ }_{12} \\
  \widehat A^{ }_{21} &  \widehat A^{ }_{22} 
 \end{array}
 \biggr)
 \biggl(
 \begin{array}{cc} 
   U & 0 \\ 
   0 & U^*  
 \end{array}
 \biggr)
 = 
 \biggl(
 \begin{array}{ll} 
  \widehat A^{ }_{11} &  \widehat A^{ }_{12} (U^*)^2 \\
  \widehat A^{ }_{21} U^2 &  \widehat A^{ }_{22} 
 \end{array}
 \biggr)
 \;.  
\ee

Apart from depending on time, the off-diagonal components play
another important role: they contain the complex phases responsible
for CP violation. Therefore, they act as sources for lepton asymmetry. 
This suggests a way to simplify the rate equations.
Indeed we can define diagonal and off-diagonal components not only 
for the coefficients, but also for the density matrix.
In particular, the rate equation for the lepton asymmetry, 
\eq\nr{summary_na}, obtains a form in which the contributions from
the diagonal and off-diagonal components of the density matrix 
are nicely separated (we denote $\nF \equiv \nF(\kT^{ })$): 
\ba
 {Y}_a' - \frac{\YB'}{3} 
 \!\!\! & = & \!\!\!  
 \frac{4}{s} \int_{\vec{k}^{ }_\rmii{$T$}} \Bigl\{ 
 \sum_{\I}
 |h^{ }_{\I a}|^2
 \Bigl[
    \widehat Q^{-}_{\I\I} \, 
    \widetilde \rho\,{}^-_{\I\I} 
 \; - \; 
   \bar \mu^{ }_a\, \widehat Q^{+}_{\I\I}\, \nF (1-\nF) 
 \; - \; 
   \bigl(
   \bar{\mu}^{ }_a \, 
        \widehat R^{+}_{\I\I} 
   + \bmuY^{ } \,
        \widehat S^{+}_{\I\I} 
   \bigr)  
   \bigl( \nF - \widetilde \rho \, {}^+_{\I\I} \bigr)
 \Bigr]
 \nn 
 & + & 
 2 \re (h^{ }_{1 a}h^*_{2 a})\,
   \widehat Q^-_{\{12\}}
  \re\bigl(  U^2 \widetilde \rho \, {}^-_{12} \bigr)
 \; - \;  
  2 \im (h^{ }_{1 a}h^*_{2 a})\,
   \widehat Q^+_{\{12\}}
 \im \bigl(  U^2 \widetilde \rho \, {}^+_{12} \bigr)
 \;\Bigr\}
 \;. \la{summary_na_2}
\ea
Here the coefficients have been evaluated up to leading order
in small chemical potentials. The terms proportional to chemical potentials 
are washout terms, the others are source terms. At early times, the solution
is dominated by the source terms on the second row. 

Consider then the source terms for the density matrix. 
The key point is that the helicity asymmetry, parametrized by 
$ \widetilde \rho \, {}^{-}_{\I\I} $, is odd in parity (P). Given that
sterile neutrinos are their own antiparticles, 
it is even in charge conjugation (C). Therefore it is odd in CP, just like
lepton asymmetries. Consequently
$ \widetilde \rho \, {}^{-}_{\I\I} $ is as small as lepton 
asymmetries, and in general much smaller than the other components 
of the density matrix. Moreover, the off-diagonal components 
$ \widetilde \rho \, {}^{\pm}_{12} $ are much smaller than the 
diagonal components $ \widetilde \rho \, {}^{+}_{\I\I} $, because both
their initial values and their equilibrium values vanish.\footnote{%
 For the benchmark point analyzed in \se\ref{se:numerics} we observe
 that the infrared (IR) modes of $ \widetilde \rho \, {}^{-}_{12} $ 
 can be as large as their $ \widetilde \rho \, {}^{+}_{\I\I} $ counterparts
 before oscillations become relevant. However,
 once oscillations have become fast and we make use of  
 the simplified equations below,  
 $ \widetilde \rho \, {}^{-}_{12} $
 is very close to its vanishing equilibrium value.} 
To summarize, we can assume that the solution satisfies 
\be
 | \bar{\mu}^{ }_a | \sim  
 | \widetilde \rho \, {}^{-}_{\I\I} | \; \ll \; 
 | \widetilde \rho \, {}^{\pm}_{12} | \; \ll \;
 | \widetilde \rho \, {}^{+}_{\I\I} |  
 \;. \la{counting}
\ee

In order to write the evolution equations in this limit, 
it is helpful to compactify the notation somewhat, denoting
\ba
 &&
 r^{ }_{12} \; \equiv \; 
 \sum_a \re (h^{ }_{1 a}h^*_{2 a})
 \;, \quad
 i^{ }_{12} \; \equiv \; 
 \sum_a \im (h^{ }_{1 a}h^*_{2 a})
 \;, \\ 
 && \widehat\Gamma^{+}_{\I} \; \equiv \; 
 2 \sum_a |h^{ }_{\I a}|^2 \widehat Q^{+}_{\I\I}
 \;, \quad
 \widehat\Gamma^{+}_\rmi{mix} \; \equiv \; 
 \frac{ \widehat\Gamma^{+}_{1} + \widehat\Gamma^{+}_{2} }{2} 
 \;, \quad
 \widehat Q^{ }_0
 \; \equiv \;
 \frac{T^2}{24\kT^{ } c_s^2 H}
 \;. 
 \hspace*{5mm} 
\ea
Now, for the diagonal helicity-symmetric 
$ \widetilde \rho \, {}^{+}_{\I\I} $, 
all terms on the right-hand side involving $\bar{\mu}^{ }_a$, 
$ \widetilde \rho \, {}^{-}_{\I\I} $, or
$ \widetilde \rho \, {}^{\pm}_{12} $, are small. Therefore
the evolution equation reads  
\be
  (\widetilde{\rho}\,{}^+_{\I\I})' =
 \widehat \Gamma^+_{\I}
 \bigl(\nF - \widetilde{\rho}\,{}^+_{\I\I} \bigr)
 \quad \mbox{(no sum over $I$)}
 \;. \la{rhop_11}
\ee
In the numerics, other terms are trivially included, and
in general they do affect the final results on a few percent
level, however eq.~\nr{rhop_11} is sufficient 
for a qualitative understanding. 

As far as the rate equations for $  \widetilde \rho \, {}^{\pm}_{12}  $
are concerned, we need to include washout contributions from 
$  \widetilde \rho \, {}^{\pm}_{12}  $ itself, 
as well as source terms from the large 
$ \widetilde{\rho}\,{}^+_{\I\I} $. 
The latter can contribute both through   
a unitary oscillation part (parametrized by $\widehat Q^{ }_0$)
as well as through
a decay/production part (parametrized by $\widehat Q^{\pm}_{12}$):
\ba
 \bigl( \widetilde \rho \, {}^+_{12} \bigr)'
  & = &  
 - \widehat \Gamma^{+}_\rmi{mix}
 \, \widetilde \rho \, {}^+_{12} 
  +  
 r^{ }_{12} (U^*)^2 
 \Bigl\{ 
   i \widehat Q^{ }_{0} 
   (\widetilde \rho \, {}^+_{22} - \widetilde \rho \, {}^+_{11})
   + \widehat Q^+_{21} (\nF - \widetilde \rho \, {}^+_{11})
   + \widehat Q^+_{12} (\nF - \widetilde \rho \, {}^+_{22})
 \Bigr\} 
 \;, \hspace*{5mm} \nn  \la{rhop_12} \\  
 \bigl( \widetilde \rho \, {}^-_{12} \bigr)'
  & = & 
 -  \widehat \Gamma^{+}_\rmi{mix}
  \, \widetilde \rho \, {}^-_{12} 
  + i^{ }_{12} (U^*)^2 
 \Bigl\{ 
    \widehat Q^{ }_0
   (\widetilde \rho \, {}^+_{22} - \widetilde \rho \, {}^+_{11})
   -i \widehat Q^-_{21} (\nF - \widetilde \rho \, {}^+_{11}) 
   -i \widehat Q^-_{12} (\nF - \widetilde \rho \, {}^+_{22})
 \Bigr\} 
 \;. \hspace*{8mm} \nn \la{rhom_12} 
\ea
The other components follow from  
$
 \widetilde \rho\, {}^+_{21} = (\widetilde \rho\, {}^{+}_{12})^*
$ 
and
$
 \widetilde \rho\, {}^-_{21} = (\widetilde \rho\, {}^{-}_{12})^* 
$.

Finally, with the same notation, 
the diagonal helicity-antisymmetric  
$\widetilde \rho \,{}^-_{\I\I}$ obeys
\ba
 && \hspace*{-1.0cm}
 (\widetilde \rho\,{}^-_{11} )'
 \; = \; 
 -  \, \widehat \Gamma^{+}_{1}  
    \widetilde \rho\,{}^-_{11} 
 \; + \; 
   2 \sum_a |h^{ }_{1 a}|^2
 \Bigl[
   \bar \mu^{ }_a\, \widehat Q^{-}_{11}\, \nF (1-\nF) 
 + \bigl(
   \bar{\mu}^{ }_a \, 
        \widehat R^{-}_{11} 
   + \bmuY^{ } \,
        \widehat S^{-}_{11} 
   \bigr)  
   \bigl( \nF - \widetilde \rho \, {}^+_{11} \bigr)
 \Bigr]
 \nn  
 && \hspace*{-1.0cm} 
  - 2 \,  r^{ }_{12} \, \Bigl[ 
   \widehat Q^+_{12}
  \re\bigl(  U^2 \widetilde \rho \, {}^-_{12} \bigr)
 \; - \; 
   \widehat Q^{ }_0 
 \im \bigl( 
   U^2 \widetilde \rho \, {}^-_{12} \bigr) \Bigr]
 \; + \;  
    2\, i^{ }_{12} \, \Bigl[ 
   \widehat Q^-_{12}
 \im \bigl(  U^2 \widetilde \rho \, {}^+_{12} \bigr)
 \; + \; 
  \widehat Q^{ }_0
 \re\bigl( 
   U^2 \widetilde \rho \, {}^+_{12} \bigr) \Bigr]
 \;,  \nn \la{rhom_11} \\
 && \hspace*{-1.0cm}
 (\widetilde \rho\,{}^-_{22} )'
 \; = \; 
  -  \widehat \Gamma^{+}_{2} \, 
    \widetilde \rho\,{}^-_{22} 
 \; + \; 
  2 \sum_a |h^{ }_{2 a}|^2
 \Bigl[
   \bar \mu^{ }_a\, \widehat Q^{-}_{22}\, \nF (1-\nF) 
 + \bigl(
   \bar{\mu}^{ }_a \, 
        \widehat R^{-}_{22} 
   + \bmuY^{ } \,
        \widehat S^{-}_{22} 
   \bigr)  
   \bigl( \nF - \widetilde \rho \, {}^+_{22} \bigr)
 \Bigr]
 \nn 
 && \hspace*{-1.0cm}
  - 2 \, r^{ }_{12} \, \Bigl[ 
   \widehat Q^+_{21}
  \re\bigl(  U^2 \widetilde \rho \, {}^-_{12} \bigr)
 \; + \; 
   \widehat Q^{ }_{0} 
 \im \bigl( 
   U^2 \widetilde \rho \, {}^-_{12} \bigr) \Bigr]
 \; + \;  
   2\, i^{ }_{12}\, \Bigl[
   \widehat Q^-_{21}
 \im \bigl(  U^2 \widetilde \rho \, {}^+_{12} \bigr)
 \; - \; 
  \widehat Q^{ }_{0} 
 \re\bigl( 
   U^2 \widetilde \rho \, {}^+_{12} \bigr) \Bigr]
 \;. \nn \la{rhom_22}
\ea
More terms are needed than before because there are many effects
of similar (small) magnitude. In fact, there is a substantial cancellation 
in the two terms proportional to $ \widehat Q^{ }_{0} $, 
which has to be properly tracked in the numerical solution.

%
\section{Treatment of fast and slow evolutions}
\la{se:fast}

%
\subsection{Outline}

There is a specific challenge with the solution of the rate equations
of \ses\ref{se:basic} and \ref{se:power}, 
namely that certain modes evolve much faster
than others. Normally, fast evolutions should be ``integrated out'', 
so that in the actual
dynamics only slow modes appear. However, a fast rate can be 
important if it leads to a new effect, absent from the
purely slow evolution. This is the case with 
sterile neutrino oscillations,  
leading to CP violation, 
and with anomalous baryon plus lepton number violation,  
converting a part of the total lepton asymmetry into a baryon asymmetry. 

More precisely, both of these rates {\em cross} the Hubble rate during
the period under consideration~\cite{as}, and therefore play 
a crucial role. At very high temperatures, 
the sterile neutrino oscillation rate is much smaller than the Hubble 
rate. Then there is no time for CP violation to take place, and no 
lepton asymmetries get generated. Around a certain temperature, 
referred to as the oscillation temperature 
$T^{ }_\rmi{osc}$ (numerically $T^{ }_\rmi{osc} \sim 10^4 \ldots 10^5$~GeV
for the benchmarks considered here), 
the oscillation rate is similar to the Hubble rate, and individual
lepton asymmetries get generated. Later on the oscillation rate
is much faster than the Hubble rate: fast oscillations can be 
averaged over, and the evolution becomes
``decoherent''. 

In contrast, the baryon plus lepton number 
violation rate starts by being much faster than the Hubble 
rate. Later on it rapidly switches off, at a temperature 
referred to as the sphaleron temperature
$T^{ }_\rmi{sph}$ (numerically $T^{ }_\rmi{sph} \sim 130$~GeV). 
For $T \ll T^{ }_\rmi{sph}$, this rate is exponentially small, and
baryon number becomes a conserved quantity. 

In the remainder of this section we show how the fast modes, 
whose direct numerical integration is challenging, can be handled. 
The basic idea for their treatment is that we solve
their equations of motion within a ``static'' background of the 
slow modes, which appear effectively as parameters in the solution. 
This solution is then inserted into the equation of
motion of the slow modes. Thereby the rate equations of 
the slow modes get modified through ``virtual'' fast corrections. 
This is similar in spirit to the usual effective theory approach.

%
\subsection{Anomalous baryon number violation}

Consider first the anomalous baryon number violation rate, discussed in 
\eqs\nr{Yap}--\nr{summary_nB}. 
Let us define $\YBpL^\rmi{eq}$ as the value of $\YBpL^{ }$ at which 
$\tmuBL^{ }$ from \eq\nr{mu_lowT}, 
and consequently $F^{ }_\rmi{diff}$ from \eq\nr{summary_nB}, 
vanishes~\cite{khleb0}: 
\be
 \YBpL^\rmi{eq} \; \equiv \; \frac{23}{79} 
 \sum_a \biggl( Y^{ }_a - \frac{\YB^{ }}{3} \biggr)
 \;. \la{nBLeq}
\ee
Then the evolution equation for $\YBpL^{ }$ 
(cf.\ \eq\nr{YBLp}) can be rewritten as 
\be
 \YBpL'  \; = \; 
 \sum_a F_a^{ } - \gamma\, (\YBpL^{ } - \YBpL^\rmi{eq}) 
 \;, \quad
 \gamma \; \equiv \;
 \frac{79 \nG^2 \Gamma^{ }_\rmi{diff}}{216 c_s^2 H T^3} 
 \;. 
\ee
Assuming that $\sum_a F^{ }_a$, 
$\YBpL^\rmi{eq}$ and $\gamma$ vary slowly,  
the fast evolution determined by 
$\gamma$ can be integrated exactly in a short time
interval $x - x^{ }_0 \ll x^{  }_0$, resulting in 
\be
 \YBpL^{ }(x)
 \; = \; 
 \YBpL^\rmi{eq} 
 + \frac{\sum_a F_a^{ }}{\gamma} 
 +  \biggl[ 
   \YBpL^{ }(x^{ }_0)
 - \YBpL^\rmi{eq}
 - \frac{\sum_a F_a^{ }}{\gamma}
    \biggr] \, e^{-\gamma\, (x- x^{ }_0)}
 \;. \la{YBL_large}
\ee
This equation applies both for $\gamma\, (x - x^{ }_0) \gg 1$ and 
$\gamma\, (x - x^{ }_0) \ll 1$. The resulting value of $\YBpL^{ }$
affects the evolution of the 
slow modes through \eqs\nr{bmua}--\nr{mu_lowT}.\footnote{%
 As an alternative recipe, leading in practice to
 indistinguishable results, we may equate $\YBpL^{ }$ with
 \eq\nr{nBLeq} down to $T\sim 140$~GeV, and 
 solve \eq\nr{YBLp} exactly at lower temperatures.  
 We also note that upon completion of our work, a paper
 appeared discussing other approaches to a treatment
 of the sphaleron rate~\cite{shintaro_new}.
 } 

%
\subsection{Fast sterile neutrino oscillations}
\la{ss:osc}

The second fast term originates from sterile neutrino 
oscillations, described by $\Hfast^{ }$, 
cf.\ \eq\nr{Hfast}. For our benchmark parameter values,  
$\Hfast^{ } \sim 10^8$ 
for $k\sim 3T$ at $T \sim T^{ }_\rmi{sph}$, 
and tracking the corresponding oscillations on par with the 
slow evolution poses a challenge.  

Let us, however, look at the fast evolution on its own, in a given
background of the slow modes. 
Consider the form of $U$ from \eq\nr{U_def}, 
{\it viz.}
\be
 U(x) \; = \; 
 \exp\Bigl\{
 i \int_{x^{ }_0}^{x}
 \! {\rm d}x' \, \Hfast^{ }(x') 
 \Bigr\}\, 
 U(x^{ }_0)
 \;. \la{U_sol}
\ee
This is not integrable because of the non-trivial
$x$-dependence of $\Hfast^{ }$. 
Suppose, however, that we integrate
only over a short period of time
(i.e.\ small interval of $x$, so that 
$(x - x^{ }_0) \partial^{ }_x \Hfast^{ }\ll \Hfast^{ } $).
Then we can expand $ \Hfast^{ } $ in slow variations, 
and to leading order use a constant $\Hfast^{ }$, 
\be 
 U(x)  \;\approx\; 
 e^{
 i   \Hfast^{ }  (x - x^{ }_0)
 } \,
 U(x^{ }_0)
 \;. \la{phase_large}
\ee
With this form, the fast oscillatory dynamics of 
\eqs\nr{rhop_12}--\nr{rhom_22} can be integrated. 

Concretely, denoting
\ba
 \mathcal{F}^{+}_{ } & \equiv & 
 r^{ }_{12} 
 \Bigl\{ 
   i \widehat Q^{ }_{0} 
   (\widetilde \rho \, {}^+_{22} - \widetilde \rho \, {}^+_{11})
   + \widehat Q^+_{21} (\nF - \widetilde \rho \, {}^+_{11})
   + \widehat Q^+_{12} (\nF - \widetilde \rho \, {}^+_{22})
 \Bigr\} 
 \;, \la{Fplus} \\ 
 \mathcal{F}^{-}_{ } & \equiv & 
 -i \times i^{ }_{12} 
 \Bigl\{ 
   i  \widehat Q^{ }_0
   (\widetilde \rho \, {}^+_{22} - \widetilde \rho \, {}^+_{11})
   + \widehat Q^-_{21} (\nF - \widetilde \rho \, {}^+_{11}) 
   + \widehat Q^-_{12} (\nF - \widetilde \rho \, {}^+_{22})
 \Bigr\} 
 \;, \la{Fminus}  
\ea
the solution for $\widetilde \rho\,{}^{\pm}_{12}$
from \eqs\nr{rhop_12} and \nr{rhom_12}, multiplied by $U^2$
as is needed in \eqs\nr{summary_na_2}, 
\nr{rhom_11} and \nr{rhom_22},  reads
\be
  U^2 \widetilde \rho \, {}^{\pm}_{12}(x)
  \; \approx \;  
 \mathcal{F}^{\pm}_{ }  \times
 \frac{e^{(2 i \Hfast^{ } - \widehat \Gamma^+_\rmi{mix})(x-x^{ }_0)} - 1}
 {2 i \Hfast^{ } - \widehat \Gamma^+_\rmi{mix}} 
 \; + \; 
  e^{(2 i \Hfast^{ } - \widehat \Gamma^+_\rmi{mix})(x-x^{ }_0)}
  \, U^2_0 \, \widetilde \rho \, {}^{\pm}_{12}(x^{ }_0) 
 \;, \hspace*{5mm} \la{rhop_12_soln} 
\ee
where $U_0^2 \equiv U^2(x^{ }_0)$. Let now $\langle \ldots \rangle^{ }$
denote an average of the solution over one oscillation period
centered around $x=\bar{x}$.
Then, to leading order in $1/\Hfast^{ }$, 
\be
 \langle 
   U^2 \widetilde \rho \, {}^{\pm}_{12}
 \rangle^{ } 
 \; = \; 
 \frac{1} {2 \Hfast^{ }} 
 \biggl[ 
 {i \mathcal{F}^{\pm}_{ }} 
 \; - \; 
 {i \Gamma^+_\rmi{mix}}
  e^{(2 i \Hfast^{ } - \widehat \Gamma^+_\rmi{mix})(\bar{x}-x^{ }_0)}
  \, U^2_0 \, \widetilde \rho \, {}^{\pm}_{12}(x^{ }_0) 
 \biggr]
 + \rmO\biggl( \frac{1}{\Hfast^{2}} \biggr)
 \;. \la{ave_UUrhopm12}
\ee
The constant part 
$ i \mathcal{F}^{\pm}_{ } / ( 2 \Hfast^{ } )$
emerges because the phase factor $(U^*)^2$ in 
\eqs\nr{rhop_12} and \nr{rhom_12} is compensated for by $U^2$
in \eqs\nr{summary_na_2}, \nr{rhom_11} and \nr{rhom_22}. 
This yields a source term for lepton asymmetries as we now show. 

In the evolution equation for the lepton asymmetries, 
\eq\nr{summary_na_2}, 
the integration over the spatial momenta 
eliminates the second 
term from \eq\nr{ave_UUrhopm12},\footnote{%
 Note that $\Hfast^{ }$ varies rapidly with $\kT^{ }$, so that 
 the integrand is oscillatory.
 } 
up to corrections of order $1/\Hfast^{2}$. 
Therefore we can replace 
\be
 \re\bigl(  U^2 \widetilde \rho \, {}^-_{12} \bigr)
 \;\longrightarrow\; 
 - \frac{\im \mathcal{F}^{-}_{ }}{ 2 \Hfast^{ }}
 \;, \quad
 \im\bigl(  U^2 \widetilde \rho \, {}^+_{12} \bigr)
 \;\longrightarrow\; 
 \frac{\re \mathcal{F}^{+}_{ }}{ 2 \Hfast^{ }}
 \;. 
\ee
Inserting $\mathcal{F}^{\pm}_{ }$
from \eqs\nr{Fplus} and \nr{Fminus} yields
\ba
 {Y}_a' - \frac{\YB'}{3} 
 \!\!\! & \approx & \!\!\!
 \frac{4}{s} \int_{\vec{k}^{ }_\rmii{$T$}} \Bigl\{ 
  \sum_{\I}
 |h^{ }_{\I a}|^2
 \Bigl[
    \widehat Q^{-}_{\I\I} \, 
    \widetilde \rho\,{}^-_{\I\I} 
 \; - \;
    \bar \mu^{ }_a\, \widehat Q^{+}_{\I\I}\, \nF (1-\nF) 
 \; - \; 
   \bigl(
   \bar{\mu}^{ }_a \, 
        \widehat R^{+}_{\I\I} 
   + \bmuY^{ } \,
        \widehat S^{+}_{\I\I} 
   \bigr)  
   \bigl( \nF - \widetilde \rho \, {}^+_{\I\I} \bigr)
 \Bigr]
 \nn 
 & + & 
 \frac{
   \re (h^{ }_{1 a}h^*_{2 a})\,  
   i^{ }_{12} \,
   \widehat Q^-_{\{12\}} 
   \widehat Q^-_{21}
    -  
   \im (h^{ }_{1 a}h^*_{2 a})\, 
   r^{ }_{12}\, 
   \widehat Q^+_{\{12\}} 
   \widehat Q^+_{21}
 }{\Hfast^{ }}\, 
 \bigl( \nF - \widetilde \rho \, {}^+_{11}  \bigr)
 \nn
 & + & 
 \frac{
   \re (h^{ }_{1 a}h^*_{2 a})\,  
   i^{ }_{12} \,
   \widehat Q^-_{\{12\}} 
   \widehat Q^-_{12}
    -  
   \im (h^{ }_{1 a}h^*_{2 a})\, 
   r^{ }_{12}\, 
   \widehat Q^+_{\{12\}} 
   \widehat Q^+_{12}
 }{\Hfast^{ }}\, 
 \bigl( \nF - \widetilde \rho \, {}^+_{22}  \bigr)
 \;\Bigr\}
 \;. \la{summary_na_3}
\ea
In this equation all terms on the right-hand side evolve slowly, 
i.e.\ without $U^2$ or $(U^*)^2$. 

We note in passing that 
if we sum over $a$, and subsequently undo the helicity
symmetrization/antisymmetrization of $\widehat Q^{\pm}_{ }$
(cf.\ \eq\nr{Q+}),  
then the numerator on the second row
of \eq\nr{summary_na_3} becomes  
\be
 r^{ }_{12} i^{ }_{12} \Bigl[ 
   \widehat Q^-_{\{12\}} 
   \widehat Q^-_{21}
    -  
   \widehat Q^+_{\{12\}} 
   \widehat Q^+_{21} \Bigr]
 = 
 -\frac{ r^{ }_{12} i^{ }_{12} }{2}
  \Bigl[ 
    \widehat Q^{ }_{(+)\{12\}} 
    \widehat Q^{ }_{(-)21}
    + 
    \widehat Q^{ }_{(+)21}
    \widehat Q^{ }_{(-)\{12\}} 
  \Bigr]
  \;,   \la{sum_na_simpl}  
\ee
and similarly for the third row. This is 
suppressed by the helicity-conserving coefficients 
$\widehat Q^{ }_{(-)} \sim M^{ }_1 M^{ }_2 / (g^2T^2)$. 
Nevertheless a total lepton asymmetry is generated 
even in the massless limit, because individual 
lepton asymmetries are generated through the source terms
in \eq\nr{summary_na_3}, and  
the washout terms (proportional to $\bar \mu^{ }_a$
in \eq\nr{summary_na_3}) depend on $a$. 

``Decoherent'' evolution equations, such as \eq\nr{summary_na_3},  
can also be obtained for the density matrix. 
If we carry out an average like in \eq\nr{ave_UUrhopm12} but for
$ \widetilde \rho \, {}^{\pm}_{12} $, a simple exercise shows that 
\be
 \langle 
   \widetilde \rho \, {}^{\pm}_{12}
 \rangle^{ } 
 \; = \; 
  e^{- \widehat \Gamma^+_\rmi{mix}(\bar{x}-x^{ }_0)}
 \biggl[ 
  \widetilde \rho \, {}^{\pm}_{12}(x^{ }_0) 
 - \frac {i   \, (U_0^*)^2 \,
  \mathcal{F}^{\pm}_{ }}{ 2 \Hfast^{ }} 
 \biggr]
 + \rmO\biggl( \frac{1}{\Hfast^{2}} \biggr)
 \;. \la{ave_rhopm12}
\ee
Therefore the average value
of
$
 \widetilde \rho \, {}^{\pm}_{12}
$
evolves slowly towards equilibrium, 
\be
 \langle 
   \widetilde \rho \, {}^{\pm}_{12}
 \rangle'
 \; \approx \; 
 - \widehat \Gamma^+_\rmi{mix} 
 \langle 
   \widetilde \rho \, {}^{\pm}_{12}
 \rangle^{ } 
 \;, 
\ee
where $\langle ... \rangle ' \equiv 
\partial^{ }_{\bar{x}} \langle ... \rangle$.

Consider finally the source terms for 
$\widetilde \rho\,{}^{-}_{\I\I}$, from the second rows of  
\eqs\nr{rhom_11} and \nr{rhom_22}. Given that 
in the end we only need the integrals over momenta of these
components, the oscillatory terms 
from \eq\nr{ave_UUrhopm12} lead to corrections suppressed
by $1/\Hfast^2$ and can again be omitted. 
Inserting the non-oscillatory parts from \eq\nr{ave_UUrhopm12} yields
\ba
 \langle\widetilde \rho\,{}^-_{11}\rangle'
 & \approx &
 - \widehat \Gamma^{+}_{1}   
 \, \langle \widetilde \rho\,{}^-_{11} \rangle
 \; + \; 
  2 \sum_a |h^{ }_{1 a}|^2
 \Bigl[
   \bar \mu^{ }_a\, \widehat Q^{-}_{11}\, \nF (1-\nF) 
 + \bigl(
   \bar{\mu}^{ }_a \, 
        \widehat R^{-}_{11} 
   + \bmuY^{ } \,
        \widehat S^{-}_{11} 
   \bigr)  
   \bigl( \nF - \widetilde \rho \, {}^+_{11} \bigr)
 \Bigr]
 \nn 
 & + & 
 \frac{r^{ }_{12}\, i^{ }_{12}}{\Hfast^{ }}
 \Bigl(  \widehat Q^{+}_{21} \widehat Q^{-}_{12} 
       - \widehat Q^{+}_{12} \widehat Q^{-}_{21} 
 \Bigr)
 \bigl( \nF - \widetilde \rho \, {}^+_{11} \bigr)
 \;,  \la{rhom_11_soln} \\[2mm]  
 \langle \widetilde \rho\,{}^-_{22} \rangle'
 & \approx &  
 - \widehat \Gamma^{+}_{2}
 \, \langle \widetilde \rho\,{}^-_{22} \rangle 
 \; + \; 
 2 \sum_a |h^{ }_{2 a}|^2
 \Bigl[
   \bar \mu^{ }_a\, \widehat Q^{-}_{22}\, \nF (1-\nF) 
 + \bigl(
   \bar{\mu}^{ }_a \, 
        \widehat R^{-}_{22} 
   + \bmuY^{ } \,
        \widehat S^{-}_{22} 
   \bigr)  
   \bigl( \nF - \widetilde \rho \, {}^+_{22} \bigr)
 \Bigr] 
 \nn 
 & + & 
 \frac{r^{ }_{12}\, i^{ }_{12}}{\Hfast^{ }}
 \Bigl(  \widehat Q^{+}_{12} \widehat Q^{-}_{21} 
       - \widehat Q^{+}_{21} \widehat Q^{-}_{12} 
 \Bigr)
 \bigl( \nF - \widetilde \rho \, {}^+_{22} \bigr)
 \;. \la{rhom_22_soln} 
\ea
Making use of the definitions of $\widehat Q^{\pm}_{ }$ from
\eq\nr{Q+} shows that 
\be
   \widehat Q^{+}_{21} \widehat Q^{-}_{12} 
 - \widehat Q^{+}_{12} \widehat Q^{-}_{21} 
  =
 \fr12
 \Bigl[ 
  \widehat Q^{ }_{(+)12} \widehat Q^{ }_{(-)21} -
  \widehat Q^{ }_{(+)21} \widehat Q^{ }_{(-)12} 
 \Bigr]  
 \;. \la{rhom_11_supp}
\ee
This structure is proportional to the helicity-conserving 
coefficients $\widehat Q^{ }_{(-)}$ and therefore suppressed
by $M^{ }_1 M^{ }_2 / (g^2T^2)$~\cite{cptheory}. In addition,
\eq\nr{rhom_11_supp} vanishes if the dependence on 
flavour indices is symmetric; this is violated only 
by soft corrections of order $(M_1^2 - M_2^2)/(g^2T^2)$~\cite{cptheory}. 
In total, we thus find a suppression 
$\sim M^{ }_1 M^{ }_2 (M_1^2 - M_2^2)/(g^4T^4 \Hfast^{ }) $
in the source terms. 

Obviously, the method presented above 
can only be used for $\Hfast^{ }\gg 1$.
Empirically, we find that it works well  
if $\Hfast^{ }\gsim 10^3 $. Note that 
$\Hfast^{ }$ depends strongly on $\kT^{ }$,  
cf.\ \eq\nr{Hfast}, 
so smaller values of $\kT^{ }$ decohere 
earlier than large values. 
Therefore 
$  {Y}_a' - {\YB'} / {3} $ should in general get a contribution
both from 
a decoherent small-$\kT^{ }$ domain according to \nr{summary_na_3}
and from 
a coherent large-$\kT^{ }$ domain according to \eq\nr{summary_na_2}.
We have verified that after the implementation of this setup, 
our results are independent of the precise value of $\Hfast^{ }$
at which we switch from the coherent to the decoherent evolution. 

%
\section{Numerical examples}
\la{se:numerics}

%
\subsection{Outline}

For the numerical solution, we start from initial conditions at which both
the sterile neutrino density matrix and all lepton asymmetries vanish, 
at some temperature $T^{ }_\rmi{max}$. For theoretical consistency,
this temperature has to be so high that sterile neutrino 
oscillations have had 
no time to take place~\cite{as}. It can therefore be determined
from \eq\nr{Hfast}, by requiring $\Hfast^{ }\ll 1$. The solution
depends on the co-moving momentum $\kT^{ }$. 
By writing $k^{ }_\rmii{$T = T^{ }_\rmii{osc}$} \equiv 
\kappa\, T^{ }_\rmi{osc}$, evaluating thermodynamic
functions at leading order in Standard Model couplings, 
and omitting the thermal mass corrections from \eq\nr{Hfast}, we get
\be
 T^{ }_\rmi{osc} \sim 7 \times 10^4 \, \mbox{GeV} 
 \, \biggl( \frac{M}{\mbox{GeV}} \, \frac{|\Delta M|}{\mbox{MeV}} 
 \, \frac{1}{\kappa} \biggr)^{1/3}
 \;, \la{Tosc}
\ee
where $M \equiv ({M^{ }_1 + M^{ }_2})/{2}$ and 
$\Delta M \equiv M^{ }_2 - M^{ }_1$. 
We choose in practice
$
 T^{ }_\rmi{max} = 10^7  \, \mbox{GeV} 
$ 
as the initial temperature, and keep track 
of momenta $\kappa \gsim 0.01$.\footnote{%
 As elaborated upon in \se\ref{se:concl}, even very small momenta
 $\kappa \lsim 0.1$ have a surprisingly large influence. 
 } 

An important aspect of the problem concerns the dependence of the
solution on $\Delta M$. With increasing $|\Delta M|$, the value of  
$\Hfast^{ }$ at the low temperature $T^{ }_\rmi{sph}$ increases, 
and therefore
the efficiency of baryon asymmetry generation, which is 
suppressed by $1/\Hfast^{ }$ at low $T$ 
(cf.\ \se\ref{ss:osc}), decreases. 
At the same time $T^{ }_\rmi{osc}$ increases 
according to \eq\nr{Tosc}, so that there is 
a longer period between $T^{ }_\rmi{osc}$ and $T^{ }_\rmi{sph}$
for the process to take place~\cite{as}. 

Another important dependence originates from the momentum $\kT^{ }$.
According to \eq\nr{Hfast}, the oscillations start earlier for the
smallest values of $\kT^{ }$, and at a given temperature are fastest
for the small-$\kT^{ }$ modes. 
At the same time, the damping 
coefficients $\widehat{Q}$, $\widehat{R}$, $\widehat{S}$  
grow rapidly with decreasing
$\kT^{ }$~(cf.\ appendix~A). This implies that the small-$\kT^{ }$ modes
both oscillate and equilibrate much faster than the large-$\kT^{ }$ modes. 

%
\subsection{Parameter choices}

As a main benchmark point we consider a case marked with $\star$ 
in \fig{4} of ref.~\cite{n3}, which lies in the middle of the viable
domain of ``scenario II'' (cf.\ \se\ref{se:intro})
according to the  
parameter scans performed in ref.~\cite{n3}. 
In the notation of appendix~B, the input parameters read
\ba
 M^{ }_1 & = & 0.7688\,\mbox{GeV}\;, \quad
 M^{ }_2 \; = \; 0.7776\,\mbox{GeV}\;, \quad
 \mbox{``inverted hierarchy''}\;,  
 \la{params1} \\ 
 z & = & 2.444 - i3.285 \;, \quad 
 \phi^{ }_1 \; = \; - 1.857 \;, \quad
 \delta \; = \; - 2.199 \;. \la{params2} 
\ea
Here $\delta$ is a Dirac-like CP-violating phase, and 
$\im z$ and $\phi^{ }_1$ are other complex phases which are
not observable in active neutrino oscillation experiments but 
enter when sterile neutrinos are considered. 
In order to consider the same physical situation 
as in ref.~\cite{n3}, 
we have inverted the signs of the complex phases, 
for reasons explained below \eq\nr{soln_n3}.
The corresponding Yukawa couplings are 
\ba
 h^{ }_{\I a} & = & 10^{-7} \times   
 \left( 
 \begin{array}{ccc} 
 \,\,
  3.522 + i 5.341 & 
 \,\,
  0.675 + i 1.090 & 
 \,\,
  0.682 - i 1.210 \\ 
 \!\!
 -5.367 + i 3.543 &
 \!\!
 -1.104 + i 0.670 & 
 \,\,
  1.227 + i 0.696
 \end{array} 
 \right) 
 \;. \hspace*{8mm}
\ea
Here, to a good approximation, 
$h^{ }_{2 a} \approx i h^{ }_{1 a} $. 
This leads to cancellations in neutrino mass formulae whereby
active neutrino mass differences can be kept at their physical
values despite largish neutrino Yukawa couplings. 

In the domain $M^{ }_{\I} \ll gT$ that we are interested in, and
restricting to temperatures $T > 100$~GeV so that processes relevant
for the ``symmetric phase'' dominate~\cite{broken}, the coefficients 
$\widehat{Q}$, $\widehat{R}$, $\widehat{S}$ only display a powerlike
mass dependence: helicity-flipping coefficients
are mass-independent, whereas helicity-conserving coefficients 
are quadratic in masses. We have evaluated the coefficients
according to ref.~\cite{cptheory}, inserting $M = 1$~GeV as
an IR regulator where needed. As an example, for 
$T = 4\times 10^4$ GeV (cf.\ \fig\ref{fig:rhoplus_rhominus})
and $\kT^{ }=3T$, the coefficients read 
\ba
 && Q^{ }_{(+)\I\J} \; = \;  5.29 \times 10^{-3}\, T 
 \;, \quad \hspace*{3mm}
 Q^{ }_{(-)\I\J} \; = \; 
 1.16 \times 10^{-3} \,
 \frac{ 
 M^{ }_{\I}M^{ }_{\J}}{T}
 \;, \\ 
 && R^{ }_{(+)\I\J} \; = \;  -1.76 \times 10^{-3}\, T 
 \;, \quad
 R^{ }_{(-)\I\J} \; = \; 
 - 
 0.37 \times 10^{-3} \,
 \frac{
 M^{ }_{\I}M^{ }_{\J}}{T}
 \;, \\ 
 && S^{ }_{(+)\I\J} \; = \;  0.87 \times 10^{-3}\, T 
 \;, \quad \hspace*{4mm}
 S^{ }_{(-)\I\J} \; = \; 
 0.04 \times 10^{-3} \,
 \frac{
 M^{ }_{\I}M^{ }_{\J}}{T}
 \;.
\ea
The equation of state is taken from ref.~\cite{crossover}
(cf.\ also ref.~\cite{dono}).\footnote{%
 The non-trivial feature of this equation of state is that 
 the heat capacity has a noticeable peak at around the electroweak 
 crossover temperature $T \approx 160$~GeV. As a result, the Universe
 spends more time in this temperature range, diluting extra energy
 density into expansion. Therefore there is relatively speaking more time
 for various production and equilibration processes to take place
 at around $T \approx 160$~GeV. We note that in principle the effect
 of sterile neutrinos should also be included in the equation of state, 
 however this results in corrections on the percent level and is
 furthermore very difficult to implement correctly, as it requires solving
 the Einstein equations simultaneously with the other ones. 
 }
The evolution equations are integrated from 
$
 T^{ }_\rmi{max} = 10^7  \, \mbox{GeV} 
$ 
down to $T^{ }_\rmi{min} = 100$~GeV, where the 
Chern-Simons diffusion rate has switched off 
and no more baryon asymmetry is being produced.

\begin{figure}[t]

\hspace*{-0.1cm}
\centerline{%
 \epsfxsize=7.8cm\epsfbox{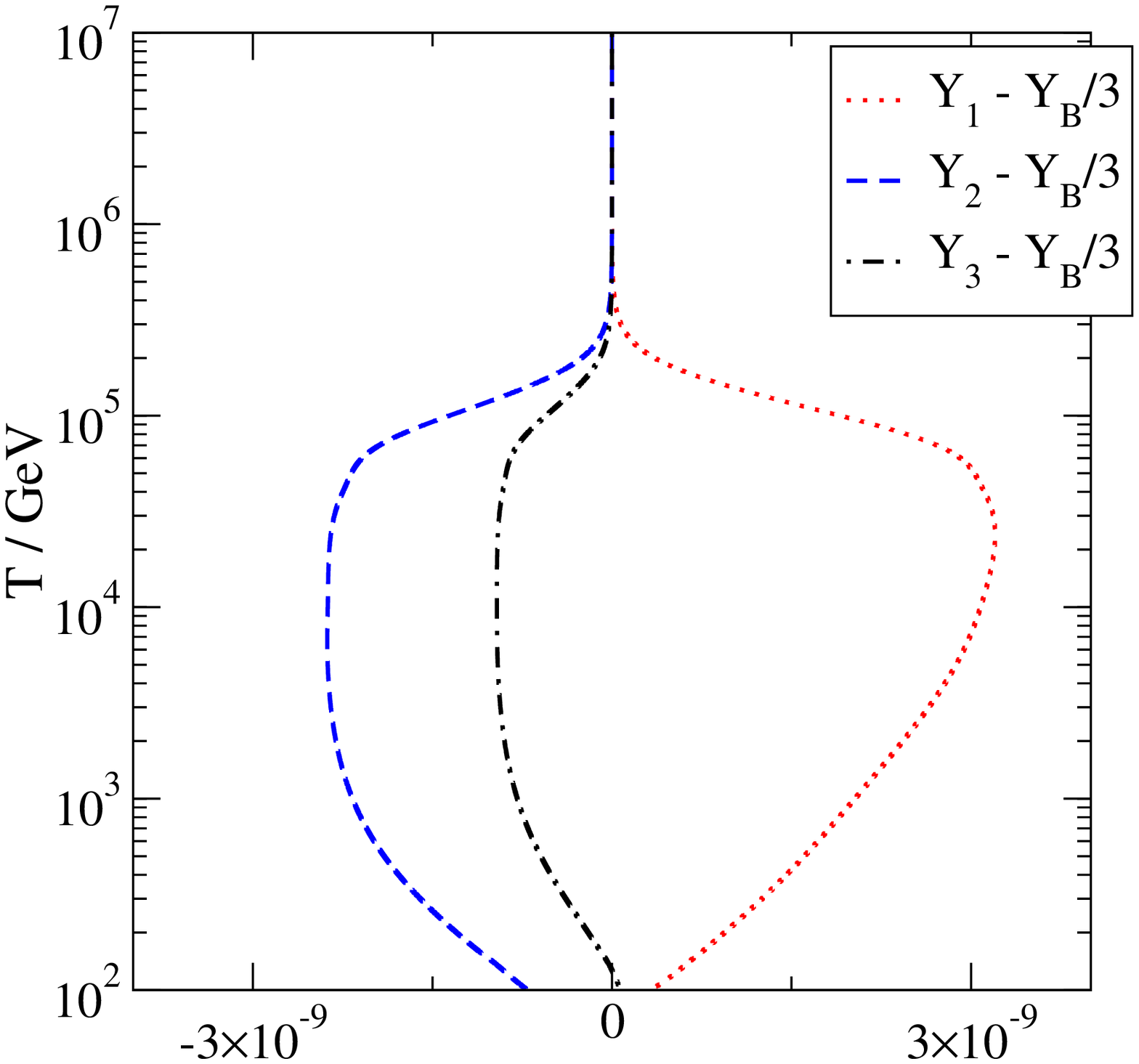}%
 \hspace{0.1cm}%
 \epsfxsize=7.5cm\epsfbox{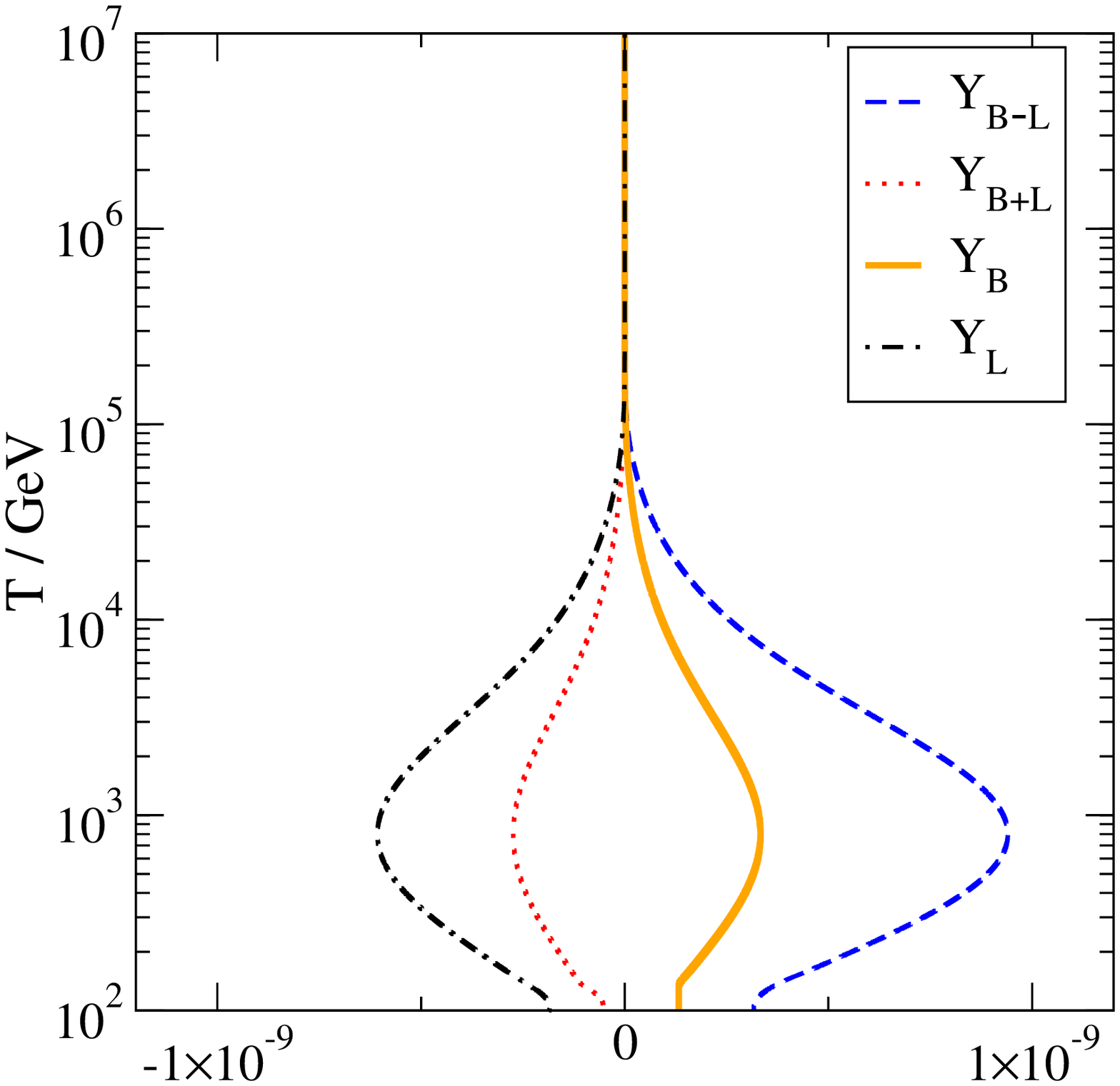}
}

\caption[a]{\small
 Left: 
 lepton minus baryon
 asymmetries $Y^{ }_a - {\YB^{ }}/{3}$ as a function of $T/\mbox{GeV}$
 for the parameters in \eqs\nr{params1}, \nr{params2}.
 Right: 
 the total baryon minus lepton 
      asymmetry $\YBmL^{ } \equiv - \sum_a (Y^{ }_a - \YB^{ }/3)$, 
 the total baryon plus lepton 
      asymmetry $\YBpL^{ }$, 
      the total baryon asymmetry $\YB^{ }$, 
      and 
      the total lepton asymmetry $\YL^{ }$.
 The baryon yield can be compared with its observed value,
 $
  \YB^{ } =
  n^{ }_\rmii{$B$} / s 
  = 0.87(1) \times 10^{-10} 
 $~\cite{planck}. 
}

\la{fig:Ya_YB}
\end{figure}

%
\subsection{Results}

\begin{figure}[t]

\hspace*{-0.1cm}
\centerline{%
 \epsfxsize=5.0cm\epsfbox{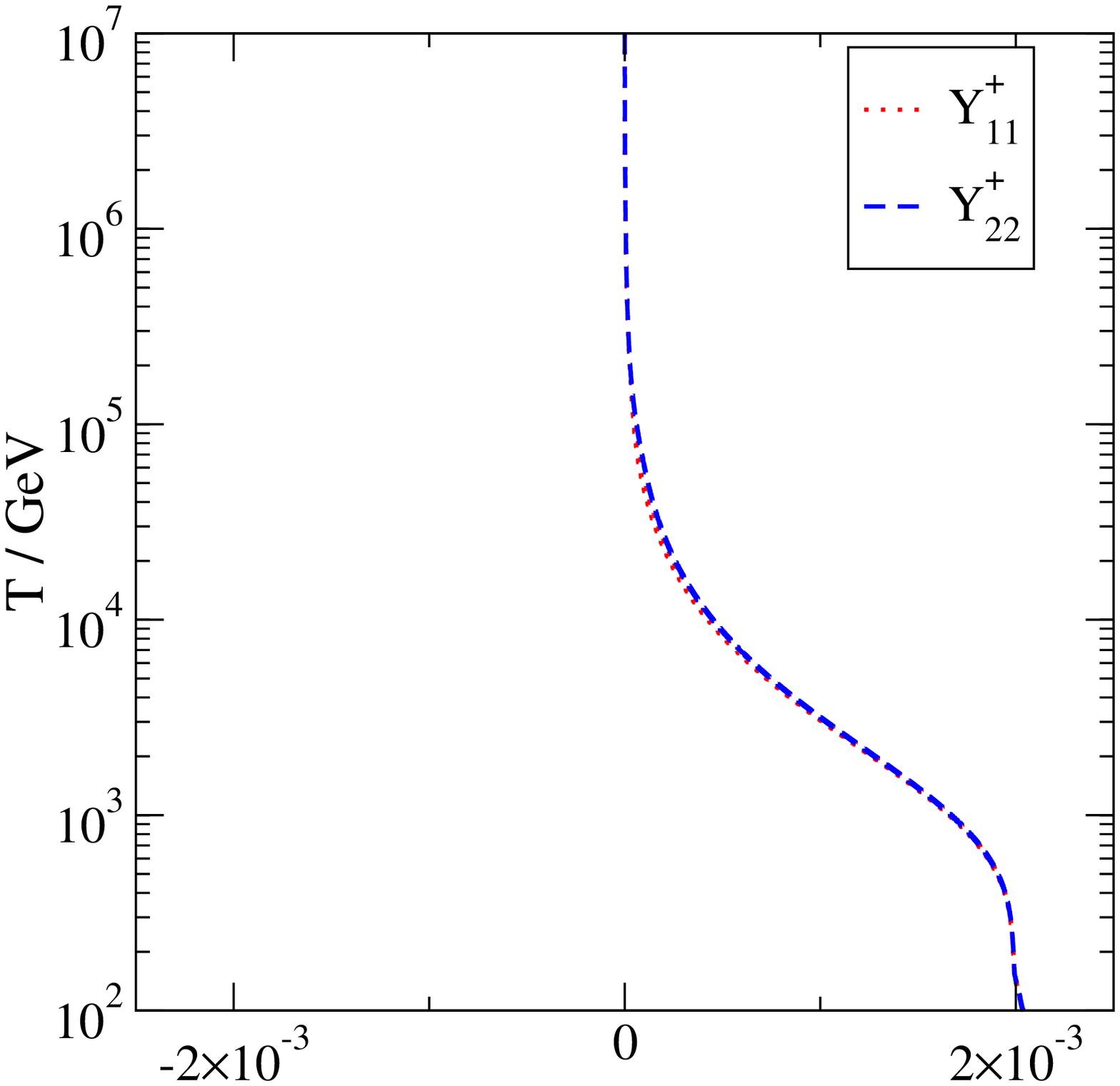}%
 \hspace{0.1cm}%
 \epsfxsize=5.0cm\epsfbox{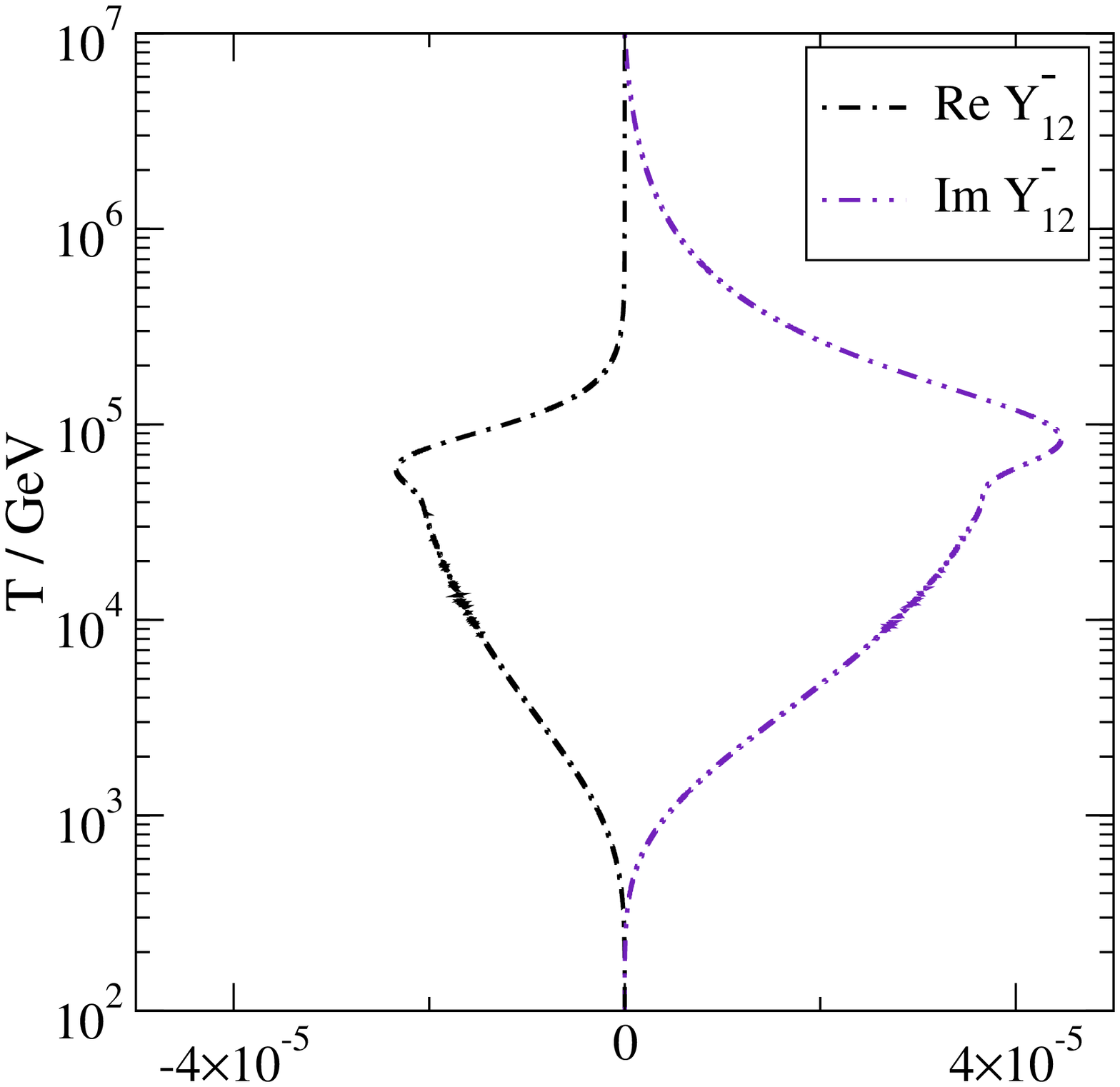}%
 \hspace{0.1cm}%
 \epsfxsize=5.0cm\epsfbox{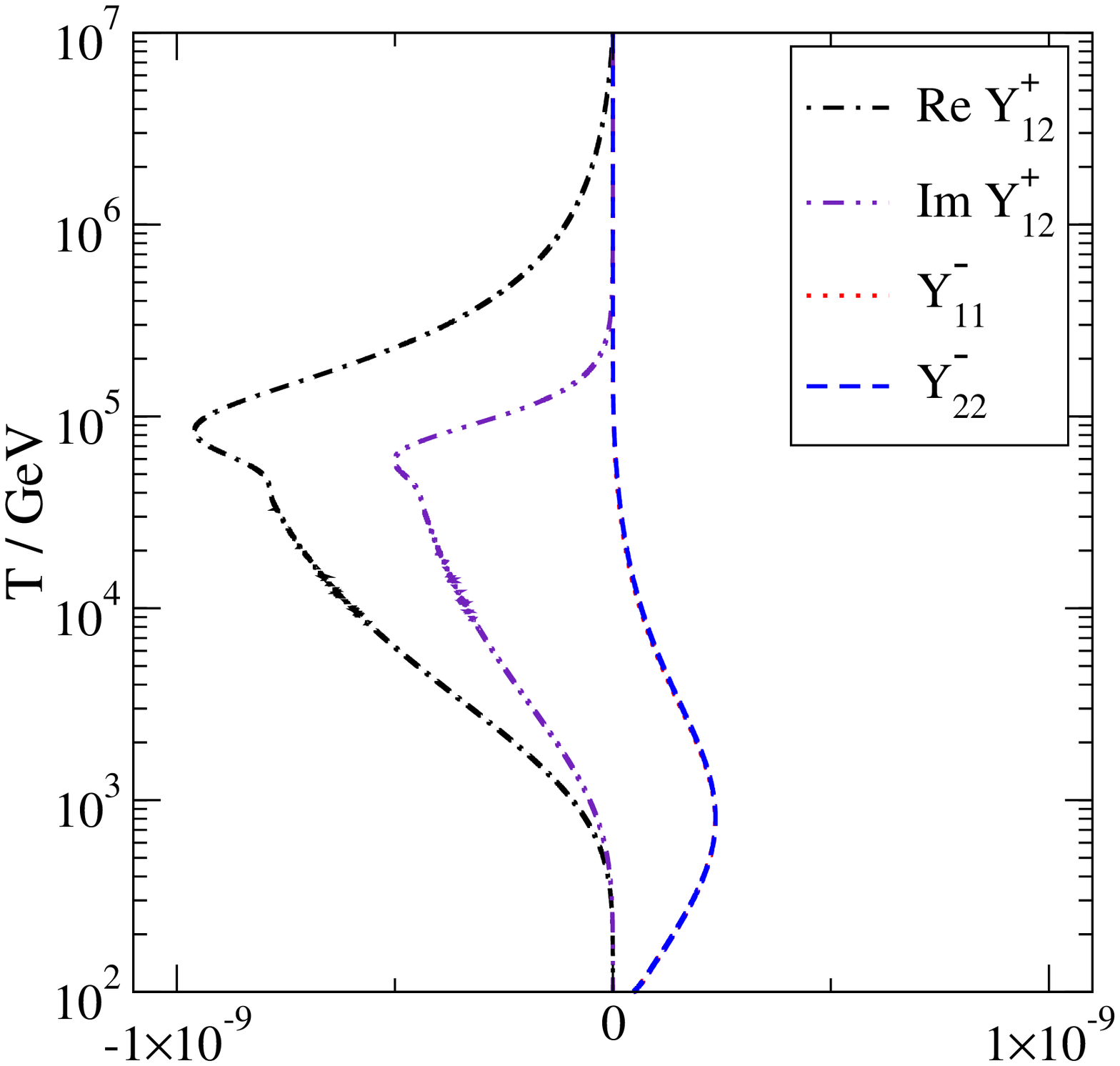}
}

\caption[a]{\small
 Averaged values of the density matrices,  
      $Y^{\pm}_{\I\J} \equiv \frac{1}{s} \int_{\vec{k}^{ }_\rmii{$T$}} 
      \widetilde \rho\, {}^{\pm}_{\I\J}$, as
 a function of $T/\mbox{GeV}$.
 Left: the helicity-symmetric diagonal components.
 Middle: the helicity-antisymmetric non-diagonal components. 
 Right: the remaining components, 
 which are of similar magnitude as the baryon asymmetry. 
}

\la{fig:Yplus_Yminus}
\end{figure}

\begin{figure}[t]

\hspace*{-0.1cm}
\centerline{%
 \epsfysize=5.5cm\epsfbox{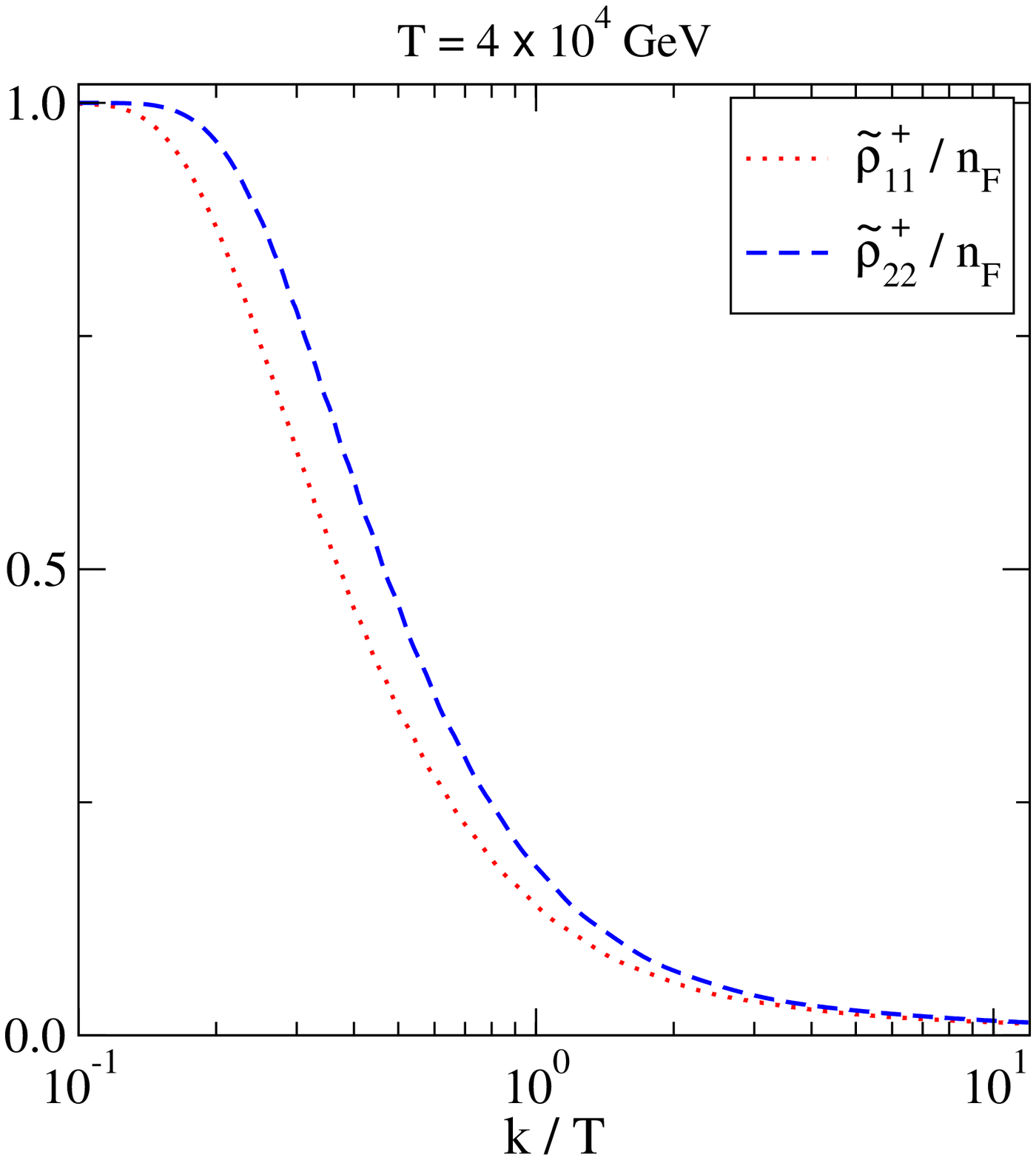}%
 \hspace{0.1cm}%
 \epsfysize=5.5cm\epsfbox{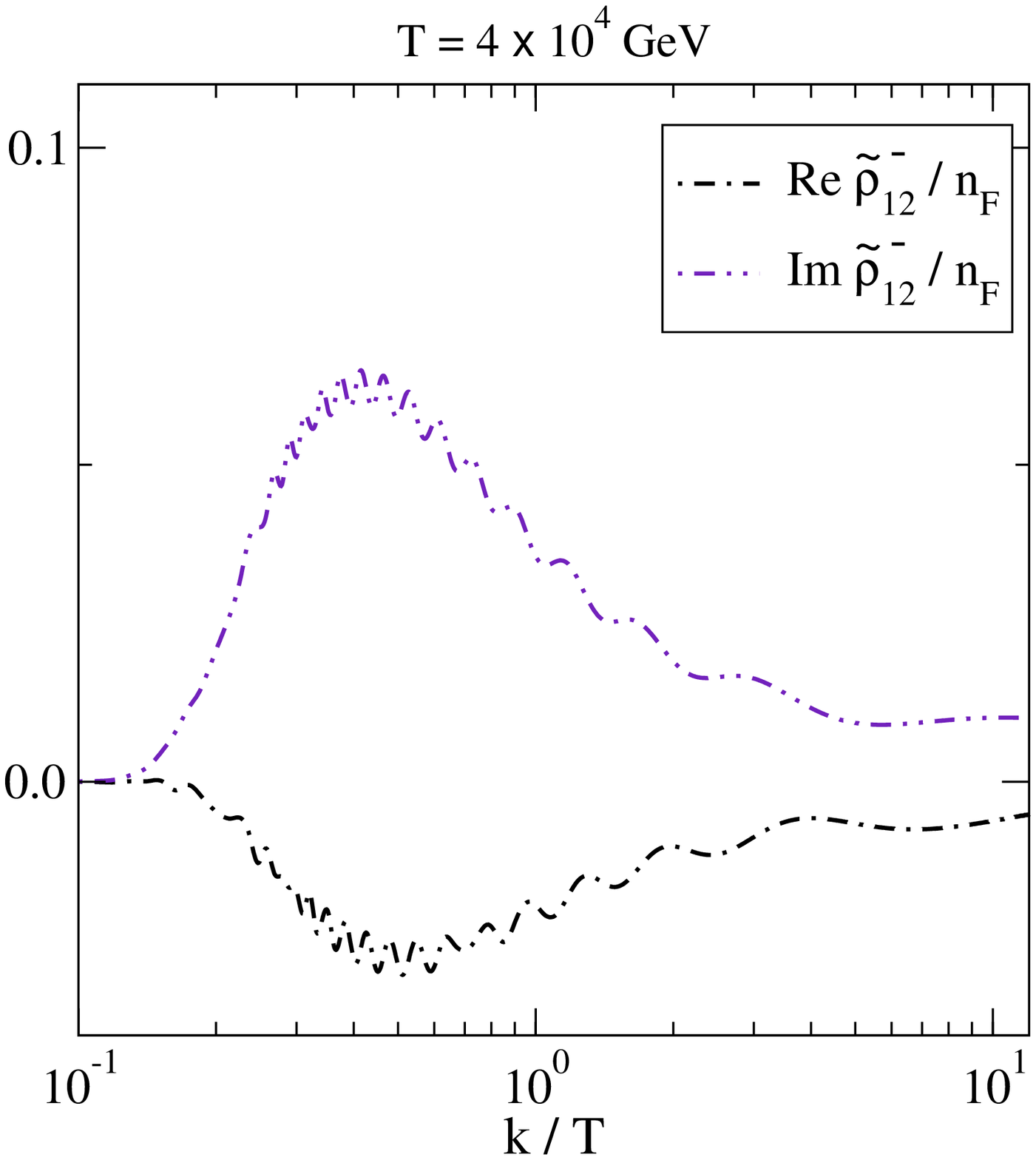}%
 \hspace{0.1cm}%
 \epsfysize=5.5cm\epsfbox{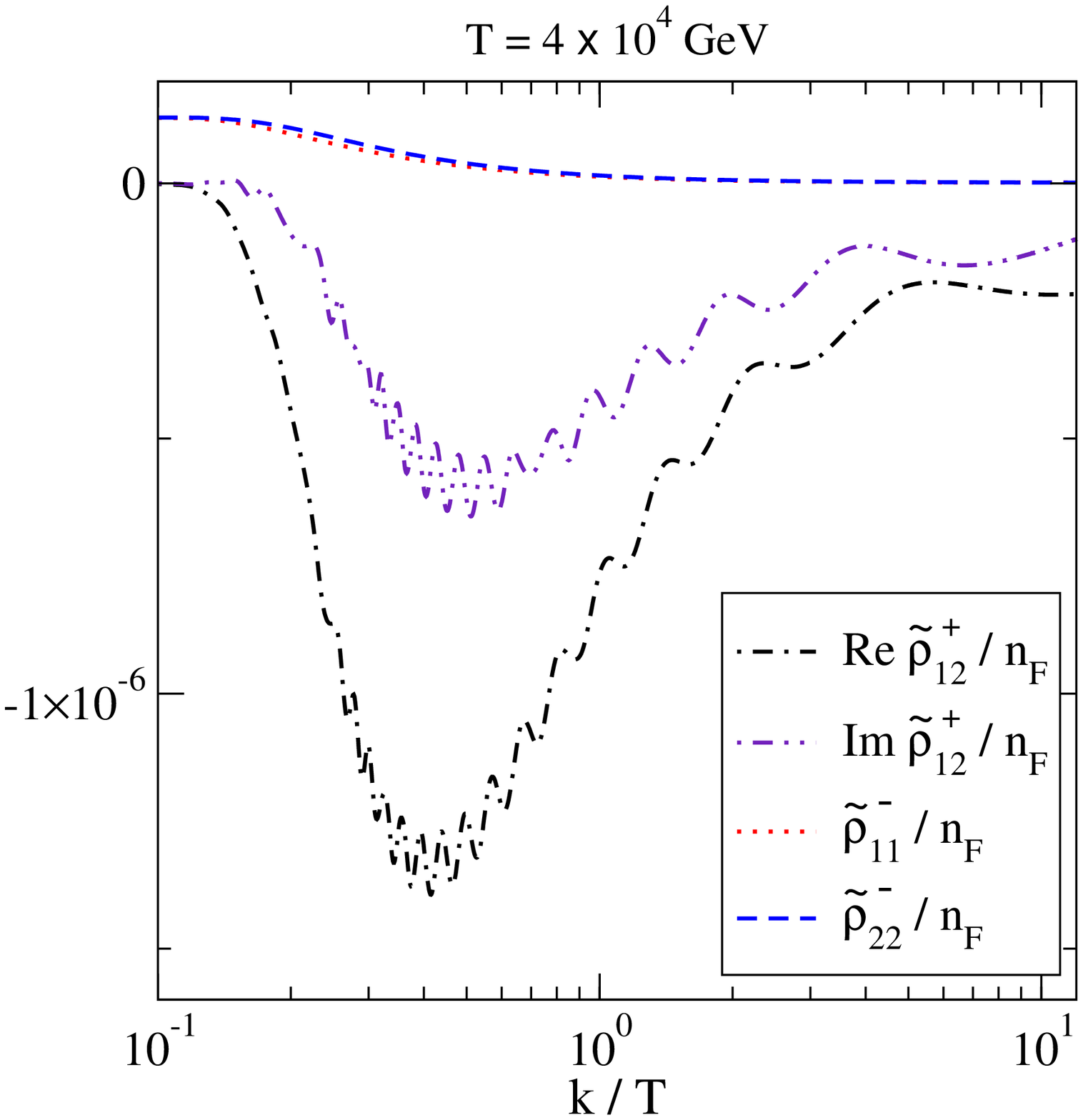}
}

\caption[a]{\small
 The shapes of various components of the density matrix
 at $T = 4\times 10^4$ GeV,
 where the production of lepton asymmetries is fastest according
 to \fig\ref{fig:Ya_YB}. The shapes have been 
 normalized to the Fermi distribution. 
 The infrared modes ($k\lsim 0.1T$) of the large components 
 $\widetilde \rho\, {}^{+}_{\I\I}$ have already reached their
 equilibrium values. 
}

\la{fig:rhoplus_rhominus}
\end{figure}

In \fig\ref{fig:Ya_YB}, 
the separate lepton minus baryon asymmetries
$Y^{ }_a - {\YB^{ }}/{3}$ are shown for the benchmark point of
\eqs\nr{params1}, \nr{params2}, together with the 
corresponding full baryon and lepton asymmetries. 
In \fig\ref{fig:Yplus_Yminus}, the integrals over
components of the density matrix are illustrated, normalised
to the entropy density. 
In \fig\ref{fig:rhoplus_rhominus}, the momentum dependence of
the density matrix is shown at 
$T \approx 4\times 10^4$~GeV, where the lepton asymmetries are being
most efficiently produced (cf.\ \fig\ref{fig:Ya_YB}). 
All shapes differ significantly from  
the Fermi distribution, with in particular the IR modes of 
$
 \widetilde \rho\, {}^{+}_{\I\I}
$ having already reached equilibrium.\footnote{%
 A similar qualitative finding was reported 
 in ref.~\cite{shintaro}, however the rate equations and coefficients
 were less complete than the current ones, for instance the rate 
 coefficients did not include the dominant
 contribution from gauge scatterings. 
 We elaborate on the significance
 of the IR modes in \se\ref{se:concl}.
 }  

Remarkably, the total baryon asymmetry that we obtain with the parameter
values of \eqs\nr{params1}, \nr{params2} is
$\YB^{ } \approx 1.3\times 10^{-10}$,  i.e.\  $\sim 50\%$
larger than the value $0.86 \times 10^{-10}$
in ref.~\cite{n3}. In other words,  
the parameter scans carried out in ref.~\cite{n3}
could be somewhat conservative in their predictions for
the viable domain. 

It can be noted from \fig\ref{fig:Ya_YB}(right)
and \fig\ref{fig:Yplus_Yminus}(right) that 
$\YLmB^{ } \equiv - \YBmL^{ }$ and 
$2 \sum_{\I} Y^{-}_{\I\I}$ cancel against each other
to a good approximation at $T \gsim 120$~GeV. This is because 
in the symmetric phase $\YLmB^{ }+ 2 \sum_{\I} Y^{-}_{\I\I}$, 
sometimes called a fermion number, 
remains zero up to corrections suppressed by 
$M^{ }_1 M^{ }_2 / (g^2T^2)$~\cite{cptheory}.
At lower temperatures the coefficient $Q^{ }_{(-)}$ kicks in
(cf.\ appendix A and ref.~\cite{es}) and fermion number violation
becomes rapidly visible.

\begin{figure}[t]

\hspace*{-0.1cm}
\centerline{%
 \epsfxsize=7.9cm\epsfbox{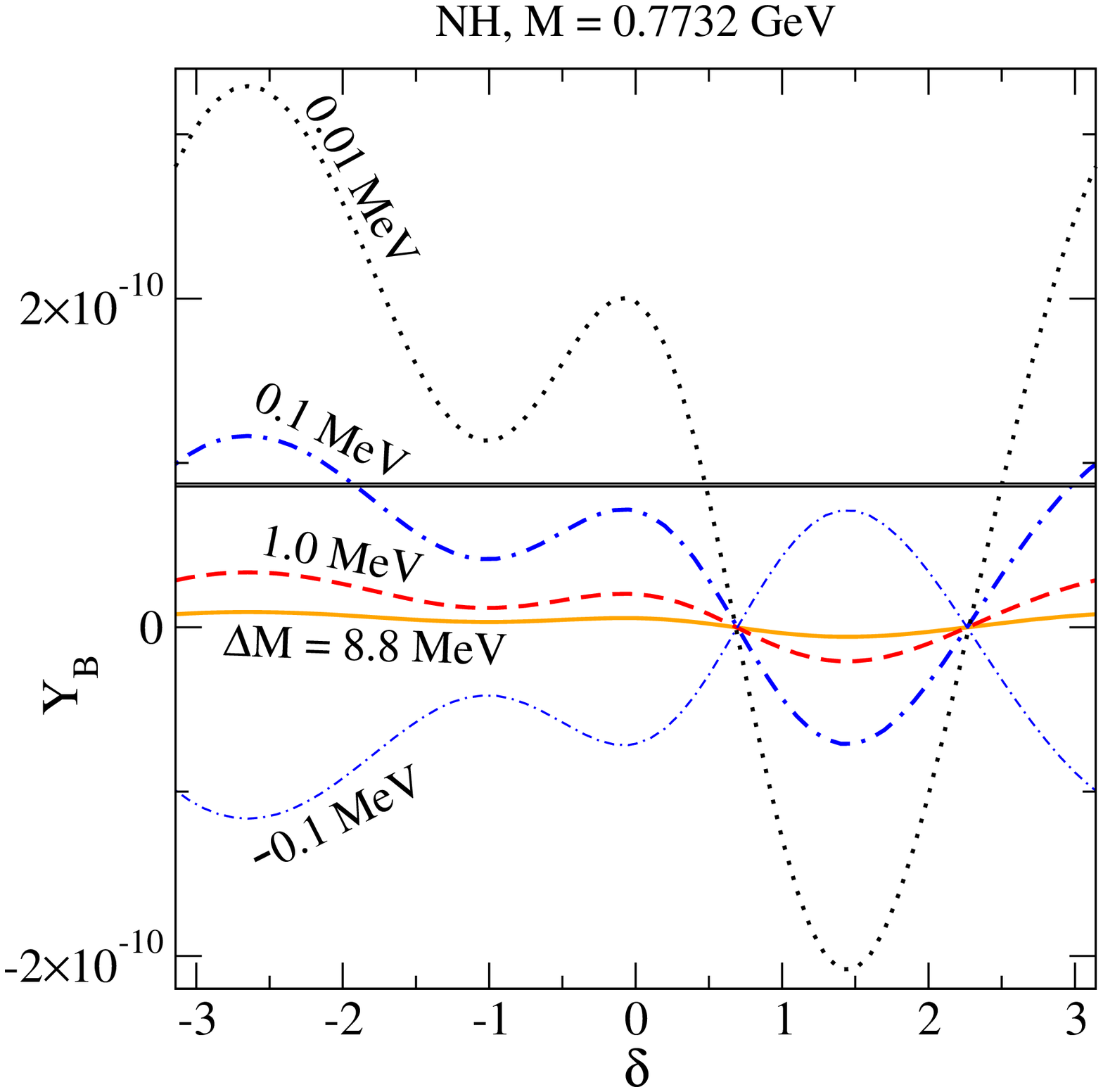}
}

\caption[a]{\small
 The dependence of the final baryon asymmetry on the sterile neutrino
 mass splitting $\Delta M$ and on the CP violating phase $\delta$. 
 The parameters are fixed according to \eqs\nr{params1}, \nr{params2}, 
 with $M \equiv (M^{ }_1 + M^{ }_2)/2$ and 
 $\Delta M \equiv M^{ }_2 - M^{ }_1$, 
 except that we now consider normal hierarchy (NH), 
 choose five different values of $\Delta M$,
 and let $\delta$ vary freely. 
 The horizontal line represents the observed value
 $
  \YB^{ } =
  n^{ }_\rmii{$B$} / s 
  = 0.87(1) \times 10^{-10} 
 $~\cite{planck}. 
}

\la{fig:scan}
\end{figure}

Finally, in \fig\ref{fig:scan}, we illustrate the dependence of the final
baryon asymmetry on the sterile neutrino mass splitting and on the CP-phase
$\delta$. The parameters have been fixed according to 
\eqs\nr{params1}, \nr{params2}, except that we now consider the
less favourable normal hierarchy. It is seen how the value of $\delta$ 
is important for obtaining the correct sign of the baryon asymmetry, 
and how the magnitude of the baryon asymmetry
is strongly affected by $\Delta M$.

%
\section{Conclusions}
\la{se:concl}

The purpose of this study has been to numerically
integrate the evolution equations derived in 
ref.~\cite{cptheory}, in order to determine how the sterile neutrino
density matrix and the lepton and baryon asymmetries evolved in 
the Early Universe. 
We find that the momentum dependence of the density matrix plays an 
important role in the solution, 
with the IR modes oscillating and equilibrating much faster than
the UV modes. Therefore the shape of the 
density matrix differs substantially
from the Fermi distribution at the time when leptogenesis
is most efficient, cf.\ \fig\ref{fig:rhoplus_rhominus}.
This effect was not included in an extensive recent parameter 
scan which otherwise employed similar rates and 
rate equations as our study~\cite{n3}.\footnote{%
 Ref.~\cite{n3} omitted
 helicity-conserving rates and terms proportional to the 
 hypercharge chemical potential, but 
 for our benchmark point
 both of these have an effect only on the 1\% level.}

As a drastic illustration for 
the importance of the momentum dependence, 
we find that even very soft modes 
$0.01T < k < 0.1T$ can give a surprisingly large
$\sim 5\%$ contribution to the final baryon asymmetry. For our
benchmark point the soft modes add up to the contribution
from the hard modes. 
Understanding more precisely the physics 
of the IR modes may merit further study.\footnote{%
 For $M,m_\phi^2 / (4T) \ll \kT^{ } \ll T$, 
 where $m_\phi^{ }$ is the thermal Higgs mass,
 the coefficient $Q^{ }_{(+)}$ grows as $\sim m_\phi^2 T/\kT^2$,
 cf.\ \eq\nr{Qplus}. 
 Inserting into \eq\nr{summary_na_3}, the contribution 
 from small $\kT^{ }$ is $\sim \int \! {\rm d} \kT^{ }/\kT^{ } 
 ( \nF^{ } - \widetilde \rho \, {}^+_{\I\I}   ) $. Therefore there
 is a logarithmic IR sensitivity, dynamically lifted if 
 the small-$\kT^{ }$ part of $ \widetilde \rho \, {}^+_{\I\I} $
 has equilibrated. 
 The IR sensitivity is even stronger in the terms influenced by
 the helicity-conserving coefficients $Q^{ }_{(-)}$, 
 cf.\ \eq\nr{Qminus}. 
 } 

As another observation from tracking the momentum dependence, 
we note that even 
though single-$\kT^{ }$ modes experience oscillations
(cf.\ \fig\ref{fig:rhoplus_rhominus} for a snapshot of spectra), 
the lepton asymmetries, which contain an integral over all momenta, 
oscillate much less (cf.\ \fig\ref{fig:Ya_YB}), 
because different momentum modes add up incoherently. 

As a benchmark point, taken from ref.~\cite{n3}, we focussed on
two sterile neutrinos which are somewhat but not 
extremely degenerate in mass, cf.\ \eqs\nr{params1}, \nr{params2}. Then
the production of lepton asymmetries is fastest at 
$T^{ }_\rmi{osc} \sim 4 \times 10^4$ GeV, much above the 
temperature $T^{ }_\rmi{sph} \sim 130$~GeV at which sphaleron
processes cease to be active. This parameter choice 
represents a typical case for the 
so-called ``scenario II'' outlined in \se\ref{se:intro}. 
For this benchmark point 
we find a baryon asymmetry $\sim 50$\% larger than 
the observed value (i.e.\ the result of ref.~\cite{n3}). 
However it would be easy to re-adjust the baryon asymmetry to the 
observed value, by modestly changing the sterile neutrino 
mass splitting or CP-violating phases, cf.\ \fig\ref{fig:scan}.

In the more restrictive ``scenario I'', which aims to generate not
only a baryon asymmetry but subsequently also much larger lepton 
asymmetries, it is natural to choose parameters leading to 
$T^{ }_\rmi{osc} \sim T^{ }_\rmi{sph}$. This case has recently
been studied in refs.~\cite{es,shintaro_new}, and we hope
to apply our methods to that situation in the future. 
Another case meriting further scrutiny is the so-called
symmetry protected scenario, $\Delta M/M\to 0$ and 
$|{\rm Im} z|\to \infty$ in the language of
\eq\nr{params2}, which leads to large neutrino Yukawa couplings 
and therefore to the best prospects for experimentally detecting 
sterile neutrinos (cf.\ ref.~\cite{new2} for an overview).

We end by remarking that our main results, 
including the rate coefficients $Q,R,S$ used, 
are publicly available from the web site
{\tt http://www.laine.itp.unibe.ch/leptogenesis/}.
We have also tabulated final results for many more benchmark 
points than discussed in this presentation. Examples of additional
points are those included in the
parameter scan illustrated in \fig\ref{fig:scan}, showing the 
dependence of the results on the sterile neutrino mass splitting
and on the CP violating parameter measurable
in active neutrino oscillation experiments. 
We would be happy to add further results on the web site, 
should readers provide us with their desired input parameters 
in the format of \eqs\nr{params1}, \nr{params2}. 

%
\section*{Acknowledgements}

We thank P.~Hern\'andez and J.~L\'opez-Pav\'on for providing us with 
the benchmark parameter values in \eqs\nr{params1}, \nr{params2}, 
for explaining some unspecified conventions of ref.~\cite{n3}, 
for confirming the issue mentioned below \eq\nr{soln_n3}, 
and for many helpful discussions.
This work was supported by the Swiss National Science Foundation
(SNF) under grant 200020-168988.

%
\appendix
\renewcommand{\thesection}{Appendix~\Alph{section}}
\renewcommand{\thesubsection}{\Alph{section}.\arabic{subsection}}
\renewcommand{\theequation}{\Alph{section}.\arabic{equation}}

%
\section{On the rate coefficients $Q$, $R$ and $S$}

The coefficients $Q$, $R$ and $S$ that parametrize 
the rate equations of \se\ref{se:basic} 
(cf.\  \eqs\nr{A}--\nr{Bminus} and \nr{Cplus}--\nr{Dminus}) 
capture the processes relevant for sterile neutrino production, 
their kinetic and chemical equilibration, as well as lepton number washout. 
They can be defined by considering the Euclidean correlator
\be
 \Pi^{ }_\rmii{E}(\tilde K) \; \equiv \; 
 \int_X e^{i \tilde K \cdot X}
 \bigl\langle 
 (\tilde{\phi}^\dagger  \ell^{ }_a)(X)  
 \, 
 (\bar{\ell}^{ }_{a} \tilde{\phi})(0) 
 \bigr\rangle
 \;, \quad \tilde K = (k^{ }_n - i \mu^{ }_a,\vec{k})
 \;, 
\ee
where $\tilde{\phi} = i \sigma_2 \phi^*$ 
is a Higgs doublet,  
$\ell^{ }_a = (\nu\, e)^T_a$ is a left-handed lepton doublet
of generation $a$, and $k^{ }_n$ is a fermionic Matsubara frequency. 
The analytic continuation $k^{ }_n - i\mu^{ }_a\to -i [k^{ }_0 + i 0^+]$ 
gives the retarded correlator $\Pi^{ }_\rmii{R}(\mathcal{K})$, 
whose imaginary part equals the spectral function 
$\rho^{ }_a (\mathcal{K})$. 
Taking matrix elements of $\rho^{ }_a (\mathcal{K})$ with on-shell
spinors leads to the desired rate coefficients: 
\ba
 \frac{
  \bar{u}^{ }_{\vec{k}\tau\J}
  \, \aL \, \rho^{ }_a(\mathcal{K}^{ }_\J) \, \aR \, 
  u^{ }_{\vec{k}\tau\I} 
  }{\sqrt{\omega^k_\I \omega^k_\J}} 
 \; \equiv \; 
    Q^{ }_{(\tau)\I\J} + \bar{\mu}^{ }_a R^{ }_{(\tau)\I\J}
  + \bmuY^{ } S^{ }_{(\tau)\I\J} + \rmO\bigl( \bar\mu^2 \bigr)
 \;, \la{QRS}
\ea
where $\mathcal{K}^{ }_\J \equiv (\omega^k_\J,\vec{k})$
with 
$
 \omega^k_\J
 \; \equiv \; \sqrt{k^2 + M_\J^2}
$;
$\aL, \aR$ are chiral projectors;
and $u^{ }_{\vec{k}\tau\I}$ is an on-shell spinor for sterile
flavour $I$ in the helicity state $\tau = \pm$. 
The lepton and hypercharge chemical potentials have been 
scaled with the temperature,   
$\bar\mu^{ }_a \equiv \mu^{ }_a/T$ and $\bmuY \equiv \muY/T$.
The specific combinations playing a role in the main body of the text are
obtained by symmetrizing or anti-symmetrizing
the original coefficients with respect to helicity,
and in some cases by symmetrizing them with respect to flavour indices:  
\be
 Q^{\pm}_{\I\J} \; \equiv \; 
 \frac{Q^{ }_{(+)\I\J} \pm Q^{ }_{(-)\I\J}}{2}
 \;, \quad
 Q^{\pm}_{\{\I\J\}} \; \equiv \; 
 \frac{Q^{\pm}_{\I\J} + Q^{\pm}_{\J\I}}{2}
 \;. 
 \la{Q+}
\ee

As an example, 
consider very high temperatures and Born level processes.
As discussed in ref.~\cite{interpolation}, 
one has to omit the lepton thermal mass $m^{ }_\ell$ from the Born computation
since it is not a mass in the usual sense but a modification of the 
dispersion relation at large momenta. 
The dominant processes are Higgs decays and inverse decays~\cite{bb1}, and
we may write 
\ba
 && \hspace*{-1.0cm} 
 \rho^{1\leftrightarrow 2}_a(\mathcal{K}^{ })
 \; = \; 
 \int_{\vec{p}} 
 \frac{ \pi 
 \, \delta( \epsilon^{ }_\phi - k^{ }_0 - p ) 
 }{2 p\, \epsilon^{ }_\phi}
 \bigl[ \nB^{ }(\epsilon^{ }_\phi - \muH^{ })
  + \nF^{ }(p + \muLa^{ }) \bigr] \, \bsl{\mathcal{P}} \,
  \la{born} \\  
 & = & 
 \int_{\frac{m_\phi^2}{4\kp} - \km}^{\frac{m_\phi^2}{4\km} - \kp} 
 \!\!\! {\rm d}p \, 
 \frac{  
  \nB^{ }(k^{ }_0 + p - \muH^{ })  
  +  
  \nF^{ }(p + \muLa^{ })
 }{16\pi k^3}
 \, \Bigl[ 
  2 p\, \bigl( k^{ }_0\, \bsl{\mathcal{K}} - M^2 \gamma^{ }_0 \bigr) 
  +  
  (m_\phi^2 - M^2) \, 
  \vec{k}\cdot \bm{\gamma}
 \Bigr] 
 \;. \nonumber
\ea
Here 
$\nB^{ }$ and $\nF^{ }$
are Bose and Fermi distributions, 
$
 \epsilon^{ }_\phi \equiv \sqrt{(\vec{p+k})^2 + m_\phi^2}
$, 
$\muH^{ } \equiv \muY^{ }/2$, 
$\muLa^{ } \equiv \mu^{ }_a - \muY^{ }/2$, 
$\mathcal{P} \equiv (p,\vec{p})$ is the lepton momentum, 
and $k^{ }_\pm \equiv (k^{ }_0 \pm k)/2$.
It is straightforward to carry out the integral over $p$, leading
to logarithms and dilogarithms. 

Taking matrix elements
according to \eq\nr{QRS} and expanding in chemical potentials 
yields the coefficients
$Q$, $R$ and $S$. For transparent expressions, let us 
restrict to $M \ll k$. Then, employing the functions
\ba
 && 
 \lnf(p) \; \equiv \; \ln \Bigl( 1 + e^{-p/T} \Bigr)
 \;, \quad
 \lif(p) \; \equiv \; \mbox{Li}^{ }_2 \Bigl(-e^{-p/T}\Bigr)
 \;, \\
 && 
 \lnb(p) \; \equiv \; \ln \Bigl( 1 - e^{-p/T} \Bigr)
 \;, \quad
 \lib(p) \; \equiv \; \mbox{Li}^{ }_2 \Bigl(e^{-p/T}\Bigr)
 \;, 
\ea 
the coefficients read
\ba
 Q^{1\leftrightarrow 2}_{(+)\I\J} & \approx &   
 \frac{m_\phi^2 T}{8\pi k^2}
 \biggl[
  \lnf\biggl( \frac{m_\phi^2}{4 k} \biggr) 
 - 
  \lnb\biggl( k + \frac{m_\phi^2}{4 k} \biggr) 
 \biggr]
 \;, \la{Qplus} \\
  Q^{1\leftrightarrow 2}_{(-)\I\J} & \approx & 
 \frac{M_{\I}^{ }M_{\J}^{ } T^2}{8\pi k^3}
 \biggl[
  \lib\biggl( k + \frac{m_\phi^2}{4 k} \biggr) 
 - 
  \lif\biggl( \frac{m_\phi^2}{4 k} \biggr) 
 \biggr]
 \;, \la{Qminus} \\ 
 R^{1\leftrightarrow 2}_{(+)\I\J} & \approx &   
 -\, \frac{m_\phi^2 T}{8\pi k^2}
  \, \nF^{ }\biggl( \frac{m_\phi^2}{4 k} \biggr) 
 \;, \la{Rplus} \\
  R^{1\leftrightarrow 2}_{(-)\I\J} & \approx & 
 -\, \frac{M_{\I}^{ }M_{\J}^{ } T^2}{8\pi k^3}
  \, \lnf\biggl( \frac{m_\phi^2}{4 k} \biggr) 
 \;, \la{Rminus} \\ 
 S^{1\leftrightarrow 2}_{(+)\I\J} & \approx &   
 \frac{m_\phi^2 T}{16\pi k^2}
 \biggl[
  \nF^{ }\biggl( \frac{m_\phi^2}{4 k} \biggr) 
  + 
  \nB^{ }\biggl( k + \frac{m_\phi^2}{4 k} \biggr) 
 \biggr]
 \;, \la{Splus} \\
  S^{1\leftrightarrow 2}_{(-)\I\J} & \approx & 
 \frac{M_{\I}^{ }M_{\J}^{ } T^2}{16\pi k^3}
 \biggl[
  \lnf\biggl( \frac{m_\phi^2}{4 k} \biggr) 
 - 
  \lnb\biggl( k + \frac{m_\phi^2}{4 k} \biggr) 
 \biggr]
 \;. \la{Sminus}
\ea
We observe that the coefficients grow rapidly at small $k$ but are then
cut off at $k \sim m_\phi^2 / (4 T)$. At the same time, they
overestimate the correct values at $k \gsim T$, because they do not contain
the lepton thermal mass $m^{ }_\ell$ that restricts the phase space in 
that region. 

In ref.~\cite{cptheory}, not only the $1\leftrightarrow 2$ 
processes but also  
$1+n \leftrightarrow 2+n$ and $2\leftrightarrow 2$ contributions to
$\rho^{ }_a(\mathcal{K})$ were included. However, in order to complete
this task, use was made of the ``collinear'' kinematic simplification 
$m_\phi^2/T, M, m^{ }_\ell, m^{ }_\phi \ll k$.
Given that we observe the domain $k \lsim m^{ }_\phi$ 
to give a significant numerical contribution to lepton asymmetries, 
we need to extrapolate the coefficients to that domain. 
In order not to grossly overestimate their values, we replace 
the collinear $1\leftrightarrow 2$ contributions
by \eqs\nr{Qplus}--\nr{Sminus} at small $k$ 
for $m^{ }_\phi > m^{ }_\ell$. We furthermore apply an overall
scaling factor to the small-$k$ corrections, in order not to inadvertently 
change the sign of the resulting coefficients in a region where their
determination is not trustworthy. 

In the domain $m^{ }_\ell < m^{ }_\phi$, i.e.\ close to the 
electroweak crossover, the small-$k$ region cannot be corrected
as above. Particularly at 120~GeV $\lsim T \lsim$ 140~GeV, there
is a lot of structure but also some numerical uncertainty in the 
determination of $Q^{ }_{(-)}$, $R^{ }_{(-)}$ and $S^{ }_{(-)}$. 
At the same time, ``indirect'' contributions, i.e.\ oscillation from
active neutrinos,  
become important at these temperatures. 
Adopting the notation and results of ref.~\cite{broken}, 
we have included them as 
$
 \delta Q^{ }_{(-)} = \im \Pi^{ }_\rmii{R} |^{ }_\rmi{indirect} / k
$, 
which indeed dominates over the direct contributions 
in the broken phase.\footnote{%
 The small-$k$ domain of the indirect contribution 
 has been investigated in ref.~\cite{miura}. 
 } 
A similar correction is expected for the chemical potential
dependence, parametrized in the symmetric phase 
by $R^{ }_{(-)}$, $S^{ }_{(-)}$, however this would
require a comprehensive re-organization of the framework, because
in the broken phase the dependence on chemical potentials is non-linear
and because the gauge potential $A^3_0$ develops an expectation value 
in addition to the hypercharge gauge potential. 
We do not dwell on these issues further here, apart from noting
that we have checked that in practice the broken phase values of  
$R^{ }_{(-)}$ and $S^{ }_{(-)}$ play very little role 
for our benchmark point. 
 
Finally we remark that 
one of the $2\leftrightarrow 2$ contributions, namely scattering
off soft Higgs bosons, was also observed to give an IR-sensitive 
contribution in ref.~\cite{cptheory}. Its \eq(3.34) needs to be refined
at $k \lsim m^{ }_\phi$, as the energy conservation constraint 
$\delta(q^{}_0-k+\sqrt{({\bf k}-{\bf q})^2+m_\phi^2})$
can only be satisfied for $k>m^{ }_\phi$ in the range $0<q^{}_0<k$.
Concretely, we now evaluate eq.~(3.34)
of ref.~\cite{cptheory} as\footnote{%
 We take the opportunity to 
 also correct a typographic error, 
 namely a missing overall factor $T/k$
 from eq.~(3.34) of ref.~\cite{cptheory}.
 This did not affect any
 numerical results presented in ref.~\cite{cptheory}. 
 } 
\ba
 \Delta S^{2\leftrightarrow 2}_{(+)} & = & 
 \frac{g_1^2 + 3 g_2^2}{(4\pi)^3 4k^2}
 \int_0^{k-m^{ }_\phi} \! {\rm d}q^{ }_0 
 \int_{k - \sqrt{(k - q^{ }_0)^2 - m_\phi^2}}
     ^{k + \sqrt{(k - q^{ }_0)^2 - m_\phi^2}} \! {\rm d}q \, 
 \frac{-T^2}{(k - q^{ }_0)^2}
 \biggl[
  \frac{k}{2} - \frac{\pi^2T^2}{2k}
 \biggr]\theta(k-m^{ }_\phi)
 \nn 
 & = & 
 \frac{(g_1^2 + 3 g_2^2)T^2}{4(4\pi)^3k}
 \biggl( \frac{\pi^2 T^2}{k^2} - 1 \biggr)
 \biggl[ 
 \ln\biggl(
   \frac{\scriptstyle \sqrt{k^2-m_\phi^2}+ k}
        {\scriptstyle m_\phi} 
 \biggr)
 -  \frac{\scriptstyle \sqrt{k^2-m_\phi^2}}
         {\scriptstyle k}
  \biggr] \theta(k-m^{ }_\phi)
 \;, 
\ea
rather than approximating the square brackets through  
$\ln(2k/m^{ }_\phi) - 1$ for all $k$. 

%
\section{Parametrization of neutrino Yukawa couplings}

We provide here a self-contained exposition of the parametrization
of neutrino Yukawa couplings, in order to be clear about our 
sign and phase conventions. 

%
\subsection{General discussion}

Let us consider the leptonic sector of a Lagrangian including right-handed
neutrinos. In order to be transparent about minus-signs, we employ
Euclidean conventions here:  
\ba
 L^{ }_E & \equiv & 
 \bar{\ell}^{ }_\rmii{L} \bsl{D}\! \ell^{ }_\rmii{L}
 + \bar{\nu}^{ }_\rmii{R} \msl{\partial}\! \nu^{ }_\rmii{R} 
 + \bar{e}^{ }_\rmii{R} \bsl{D}\! e^{ }_\rmii{R} 
 +  \frac{1}{2}
 \bigl(
 \bar{\nu}_\rmii{R}^c M^{ }_{ } \nu^{ }_\rmii{R} 
 + 
 \bar{\nu}_\rmii{R}^{ }\, M^{\dagger}_{ } \nu^c_\rmii{R} 
 \bigr)
 \nn[0.5mm] 
 & + & 
  \phi^\dagger \bar{e}^{ }_\rmii{R}\,  h_{e}^{ } \ell^{ }_\rmii{L}
 + 
  \bar{\ell}^{ }_\rmii{L}\, h_{e}^\dagger \, e^{ }_\rmii{R} \, \phi
 + 
 \tilde\phi^\dagger \bar{\nu}^{ }_\rmii{R}\,  h_{\nu}^{ } \ell^{ }_\rmii{L}
 + 
 \bar{\ell}^{ }_\rmii{L}\, h_{\nu}^\dagger \, \nu^{ }_\rmii{R} \, \tilde\phi
 \;. \la{SM_leptonic}
\ea
Here 
$\ell^{ }_\rmii{L} \equiv (\nu^{ }_\rmii{L}\, e^{ }_\rmii{L})^T$; 
$\tilde{\phi} \equiv i \sigma_2 \phi^*$ is a Higgs doublet; 
$
 \nu^c_\rmii{R} \equiv C \bar{\nu}^T_\rmii{R} 
$
denotes a charge-conjugated spinor; 
and $M$, $h^{ }_e$ and $h^{ }_\nu$ are complex
matrices with generation indices. 

Given that 
$ 
 \bar{\nu}_\rmii{R}^c M^{ }_{ } \nu^{ }_\rmii{R} = 
 \bar{\nu}_\rmii{R}^c M^{T}_{ } \nu^{ }_\rmii{R}
$, the mass matrix $M$ is symmetric, $M^T = M$. 
Through the so-called Takagi factorization
(a special case of singular value decomposition), 
it can be written as $M = V \Delta^{ }_M\, V^T$, where $V$ is
unitary and $\Delta^{ }_{M}$ 
is a diagonal matrix with real non-negative entries.
The matrices $V$ and $\Delta^2_M$ can
be found by diagonalizing the Hermitean matrix $M M^\dagger$.
Subsequently $V$ can be eliminated through a unitary rotation 
of~$\nu^{ }_\rmii{R}$. In the following we assume that this field
redefinition has been carried out, and that therefore
$M = \mathop{\mbox{diag}}(M^{ }_1, M^{ }_2, M^{ }_3)$, 
where $M^{ }_\rmii{$I$} \ge 0$
are referred to as the Majorana masses. 

The Yukawa matrix $h^{ }_e$ can also be assumed to be real and diagonal. 
Indeed, a biunitary transformation permits us to write it as 
$h^{ }_e = W_\rmii{R}^\dagger\, \Delta^{ }_{h^{ }_e} W^{ }_\rmii{L}$, 
where $W^{ }_{\rmii{R,L}}$ are unitary matrices. There is no unique
choice for $W^{ }_{\rmii{R,L}}$, but possibilities can be found by
diagonalizing the Hermitean matrices 
$\displaystyle h_{e}^{ } h_{e}^\dagger$ and 
$\displaystyle h_{e}^{\dagger} h_{e}^{ }$, 
respectively. In the following, we assume that $\ell^{ }_\rmii{L}$
has subsequently been rotated as 
$ 
 \ell^{ }_\rmii{L} \to W^{\dagger}_{\rmii{L}}\ell^{ }_\rmii{L}$ 
and $e^{ }_\rmii{R}$ as 
$ 
 e^{ }_\rmii{R} \to W^{\dagger}_{\rmii{R}} e^{ }_\rmii{R}$, 
so that $h^{ }_e$ is diagonal, 
with real positive entries proportional
to charged lepton masses. 

After the field redefinitions of $\nu^{ }_\rmii{R}$
and $\ell^{ }_\rmii{L}$, the matrix $h^{ }_\nu$ is in general
complex and non-diagonal. There are three free phases in 
$W^{ }_{\rmii{L}}$ which can be used to remove redundancies. 
Therefore, the total number of parameters 
introduced by $N=3$ right-handed neutrinos is 18 
($N$ from $M^{ }_\rmii{$I$}$ and $2N^2 - N$ from the 
complex matrix $h^{ }_{\nu}$ with three unphysical phases projected away).
Of these, 5 are currently known (two active neutrino mass differences
and three mixing angles) and 2 are frequently considered accessible 
(absolute mass scale of active neutrino masses and ``Dirac-like''
CP-violating phase in the active neutrino mixing matrix). 
The remaining 11 can be chosen as the three 3 Majorana 
masses $M^{ }_\rmii{$I$}$, 
2 ``Majorana-like'' phases in the active neutrino mixing matrix (see below), 
and 3 complex angles related to the so-called $R$ matrix of 
the Casas-Ibarra parametrization~\cite{ci} (see below). 
Combinations of these can possibly 
be constrained by $0\nu\beta\beta$ and $B$-factory-type
experiments.

As a next step, 
let us go to the Higgs vacuum, setting 
$\tilde{\phi} \simeq (v/\sqrt{2},0)^T$ where $v\simeq 246$~GeV. 
We denote
\be
 M^{ }_\rmii{D} \; \equiv \; \frac{ h_\nu^\dagger v }{ \sqrt{2} } 
 \; = \; \frac{ Y v }{ \sqrt{2} }
 \;, \la{def}
\ee 
where $Y$ corresponds to the notation of ref.~\cite{n3}. 
Then, from \eq\nr{SM_leptonic} and recalling the transformation
carried out with $M$, the mass terms in the neutrino sector read
\be
 \delta L^{ }_E = 
 \frac{1}{2}
 \bigl(
 \bar{\nu}_\rmii{R}^c M^{ }_{ } \nu^{ }_\rmii{R} 
 + 
 \bar{\nu}_\rmii{R}^{ }\, M^{ }_{ } \nu^c_\rmii{R} 
 \bigr)
 + 
 \bar{\nu}_\rmii{R}^{ } M^{\dagger}_\rmii{D}\, \nu^{ }_\rmii{L} 
 +  
 \bar{\nu}_\rmii{L}^{ } M^{ }_\rmii{D} \nu^{ }_\rmii{R} 
 \;. 
\ee
Inserting $-1 = C C$ and noting that 
$
 \nu^{T}_\rmii{R} C = 
 \bar{\nu}_\rmii{R}^c 
$, 
we can write
\be
 \bar{\nu}_\rmii{L}^{ } M^{ }_\rmii{D} \nu^{ }_\rmii{R} 
 = 
 \fr12 \bigl( 
  \bar{\nu}_\rmii{L}^{ } M^{ }_\rmii{D} \nu^{ }_\rmii{R} 
 - 
  \nu^{T}_\rmii{R}  M^{T}_\rmii{D} \bar{\nu}_\rmii{L}^{T} 
 \bigr)
 = 
 \fr12 \bigl( 
 \bar{\nu}_\rmii{L}^{ } M^{ }_\rmii{D} \nu^{ }_\rmii{R} 
 + 
 \bar{\nu}^{c}_\rmii{R} M^{T}_\rmii{D} \nu^{c}_\rmii{L}
 \bigr)
 \;, \la{master}
\ee
and similarly 
$
 \bar{\nu}_\rmii{R}^{ } M^{\dagger}_\rmii{D}\, \nu^{ }_\rmii{L}
 = 
 \bar{\nu}_\rmii{L}^{c} M^{*}_\rmii{D}\, \nu^{c}_\rmii{R}
$. 
Thereby
\be
 \delta L^{ }_E = 
 \frac{1}{2}
 \bigl(
 \bar{\nu}_\rmii{L}^{c}\; 
 \bar{\nu}_\rmii{R}^{ }
 \bigr)
 \underbrace{
 \left(
 \begin{array}{cc}
  0 & M^{*}_\rmii{D} \\ 
  M^{\dagger}_\rmii{D} & M 
 \end{array} 
 \right)}_{ \equiv\, \mathcal{M}^{ }_\nu }
 \left( 
 \begin{array}{c}
   \nu^{ }_\rmii{L} \\ 
   \nu^{c}_\rmii{R}
 \end{array}
 \right)
 + 
 \frac{1}{2}
 \bigl(
 \bar{\nu}_\rmii{L}^{ }\; 
 \bar{\nu}_\rmii{R}^c
 \bigr)
 \left(
 \begin{array}{cc}
  0 & M^{ }_\rmii{D} \\ 
  M^{T}_\rmii{D} & M 
 \end{array} 
 \right)
 \left( 
 \begin{array}{c}
   \nu^{c}_\rmii{L} \\ 
   \nu^{ }_\rmii{R}
 \end{array}
 \right)
 \;. \la{Mnu}
\ee
Here $ \mathcal{M}^{ }_\nu $ corresponds to the notation 
of ref.~\cite{n0}, representing a matrix multiplying 
$(\nu^{ }_\rmii{L}\,{\nu}_\rmii{R}^c)^T$.

The matrix $ \mathcal{M}^{ }_\nu $ is symmetric and can again 
be represented via the Takagi factorization:
\be
 \mathcal{M}^{ }_\nu = U^*\, \mbox{diag}(m^{ }_\nu, M^{ }_h)\, U^\dagger
 \;, 
\ee
where $m^{ }_\nu$ and $M^{ }_h$ are real matrices containing the active
and sterile neutrino masses, respectively. According to \eq(2.17) of 
ref.~\cite{n0}, in the seesaw limit we can write
\be
 U \approx
 \left( 
  \begin{array}{cc}
    U^{ }_\rmii{PMNS} &
      i\, U^{ }_\rmii{PMNS}\, m^{1/2}_\nu R^\dagger\, M^{-1/2} \\ 
    i M^{-1/2}\, R\, m_\nu^{1/2} &
      \mathbbm{1}  
  \end{array}
 \right)
 \;, \la{U}
\ee
where $U^{ }_\rmii{PMNS}$
is the Pontecorvo-Maki-Nakagawa-Sakata matrix, and
$R$ is orthogonal.  

As the final step, $U^{ }_\rmii{PMNS}$ can be rotated
away from active neutrino masses through 
$\nu^{ }_\rmii{L} \to U^{ }_\rmii{PMNS} \nu^{ }_\rmii{L}$. 
It is then
re-introduced into the non-diagonal parts of the weak interaction term,  
\be
 L^{ }_E \; \approx \;
 \bigl(
 \bar{\nu}^{ }_\rmii{L} U^{\dagger}_\rmii{PMNS} \,,\; 
 \bar{e}^{ }_\rmii{L} 
 \bigr)
 \bsl{D}\! 
 \biggl( 
 \begin{array}{c} 
   U^{ }_\rmii{PMNS} \nu^{ }_\rmii{L} \\ 
   e^{ }_\rmii{L} 
 \end{array}
 \biggr) 
 + \bar{e}^{ }_\rmii{R} \bsl{D}\! e^{ }_\rmii{R} 
 + \sum_{a = e,\mu,\tau } m^{ }_{a} \bar{e}^{ }_a e^{ }_a
 + \frac{1}{2} \Bigl( 
 \bar{\nu}^c_\rmii{L} m^{ }_\nu
 \nu^{ }_\rmii{L}
 + 
 \bar{\nu}^{ }_\rmii{L} m^{ }_\nu
 \nu^c_\rmii{L}
 \Bigr) 
 \;, 
\ee
where $m^{ }_{a}$ are the charged lepton masses. 
Because of the freedom of $N$ phase rotations of $W^{ }_\rmii{L}$
mentioned above, 
$U^{ }_\rmii{PMNS}$ has $N^2 - N = 6$ free parameters (see below). 

The relation in \eq\nr{U} underlies the so-called
Casas-Ibarra parametrization~\cite{ci} and its 
generalization beyond the seesaw limit~\cite{n0}. Specifically, 
combining \eqs\nr{Mnu}--\nr{U}, inspecting the upper right block, 
and expanding to leading order in $1/M$, we obtain 
\be
 M^{*}_\rmii{D} = - i 
  U^{*}_\rmii{PMNS}
  \,
 \sqrt{m^{ }_\nu}
 \, 
 R^T
 \, 
 \sqrt{M}
 \;, \quad
 M^{ }_\rmii{D} = i 
  U^{ }_\rmii{PMNS}
  \,
 \sqrt{m^{ }_\nu}
 \, 
 R^\dagger
 \, 
 \sqrt{M}
 \;.  \la{soln_n3}
\ee
We note that \eq(2.5) of ref.~\cite{n3} cites the left version
for $M^{ }_\rmii{D}$, so in comparisons with ref.~\cite{n3} we need
to flip the signs of complex phases, if we want to study the same
physical situation. 

%
\subsection{Parametrization of $U^{ }_\rmii{PMNS}$}

We proceed to the parametrization of $U^{ }_\rmii{PMNS}$, which
fixes the neutrino Yukawa couplings according
to \eqs\nr{def} and \nr{soln_n3}. 
As mentioned
above, 6 real parameters are needed: 4 ``Dirac-like'' parameters like for the 
Cabibbo-Kobayashi-Maskawa
matrix, and two additional parameters, which can be chosen as 
``Majorana-like'' phases. Ref.~\cite{n3} writes 
\be
 U^{ }_\rmii{PMNS} = 
 V^{ }_\rmii{PMNS}
 \,
 \left( 
 \begin{array}{ccc} 
  1 & 0 & 0 \\ 
  0 & e^{i \phi^{ }_1} & 0 \\
  0 & 0 & e^{i \phi^{ }_2}
 \end{array}
 \right) 
 \;, \la{phase}
\ee
where $V^{ }_\rmii{PMNS}$ is the Dirac-like part. 
The Dirac-like part is conventionally expressed as
\be
 V^{ }_\rmii{PMNS} 
 = 
 \left( 
 \begin{array}{ccc}
  c^{ }_{12} c^{ }_{13} & 
  s^{ }_{12} c^{ }_{13} & 
  \;\, s^{ }_{13} e^{-i \delta}
  \\
 - s^{ }_{12} c^{ }_{23} - c^{ }_{12} s^{ }_{13} s^{ }_{23} e^{i \delta} & 
 \;\;  c^{ }_{12} c^{ }_{23} - s^{ }_{12} s^{ }_{13} s^{ }_{23} e^{i \delta} & 
   c^{ }_{13} s^{ }_{23}
  \\ 
 \;\;  s^{ }_{12} s^{ }_{23} - c^{ }_{12} s^{ }_{13} c^{ }_{23} e^{i \delta} & 
 - c^{ }_{12} s^{ }_{23} - s^{ }_{12} s^{ }_{13} c^{ }_{23} e^{i \delta} & 
   c^{ }_{13} c^{ }_{23} 
 \end{array}
 \right)
 \;, 
\ee 
where $c^{ }_{ij} \equiv \cos\theta^{ }_{ij}$ 
and $s^{ }_{ij} \equiv \sin\theta^{ }_{ij}$.  
For the mass differences, we denote $\Delta m_{ij}^2 \equiv m_i^2 - m_j^2$.
Two cases are considered, normal hierarchy (NH) and inverted hierarchy (IH). 
According to ref.~\cite{gg}, the best-fit values are 
\ba
 \mbox{(NH)}: &&
 \theta^{ }_{12} = {33.48^\circ}^{+0.78}_{-0.75}\;, \quad
 \theta^{ }_{23} = {42.3^\circ}^{+3.0}_{-1.6}\;, \quad
 \theta^{ }_{13} = {8.50^\circ}^{+0.20}_{-0.21}\;, \\ 
 && \Delta m_{21}^2 = 7.50^{+0.19}_{-0.17} \times 10^{-5}\mbox{eV}^2 \;, \quad
    \Delta m_{31}^2 = 2.457^{+0.047}_{-0.047} \times 10^{-3}\mbox{eV}^2 
 \;,\\[2mm]
 \mbox{(IH)}: &&
 \theta^{ }_{12} = {33.48^\circ}^{+0.78}_{-0.75}\;, \quad
 \theta^{ }_{23} = {49.5^\circ}^{+1.5}_{-2.2}\;, \quad
 \theta^{ }_{13} = {8.51^\circ}^{+0.20}_{-0.21}\;, \\ 
 && \Delta m_{21}^2 = 7.50^{+0.19}_{-0.17} \times 10^{-5}\mbox{eV}^2 \;, \quad
    \Delta m_{23}^2 = 2.449^{+0.048}_{-0.047} \times 10^{-3}\mbox{eV}^2 \;.    
\ea

%
\subsection{Specialization to two sterile generations}

After the general discussion above,
we now focus on a special case. 
In the so-called $\nu$MSM parameter corner
(scenarios I and II in the language of \se\ref{se:intro}), 
one Majorana mass is very small ($\sim$ keV), and the 
corresponding Yukawa couplings are tiny, so that 
the contribution from these Yukawas to active neutrino masses 
is vanishing.  Following ref.~\cite{n3}, the small Majorana
mass is denoted by $M^{ }_3$,  
and we set $(h^{ }_{\nu})^{ }_{3a} \to 0$.\footnote{%
 In the line of work reviewed in ref.~\cite{canetti}, 
 it is rather the Majorana generation $I=1$ that is decoupled. 
 }
Consequently, the smallest of the active neutrino masses 
necessarily vanishes. Therefore active neutrino masses are
now fixed:
\ba
 \mbox{(NH)}: && m^{ }_\nu = \mathop{\mbox{diag}}
 \bigl( 0, \sqrt{\Delta m_{21}^2}, \sqrt{\Delta m_{31}^2 } \bigr)
 \;, \la{mnu_nh} \\ 
 \mbox{(IH)}: && m^{ }_\nu = \mathop{\mbox{diag}}
 \bigl( \sqrt{\Delta m_{23}^2 - \Delta m_{21}^2},
 \sqrt{\Delta m_{23}^2 },0 \bigr)
 \;. \la{mnu_ih}
\ea

After the choice $(h^{ }_{\nu})^{ }_{3a} \to 0$, 
$M^{ }_\rmii{D}$ in \eq\nr{soln_n3}
is effectively a $3\times 2$ matrix, whereas $M$ and $R$ 
are effectively $2\times 2$ matrices.
Concretely, we write 
\be
 M = \left( 
 \begin{array}{cc}
  M^{ }_1 & 0 \\ 
  0 & M^{ }_2
 \end{array}
 \right)
 \;, \quad
 R = 
 \left( 
 \begin{array}{cc}
  \;\, \cos z & \sin z \\ 
  - \sin z & \cos z
 \end{array}
 \right)
 \;, \quad z \in \mathbbm{C}
 \;.
\ee
Eqs.~\nr{def} and \nr{soln_n3} imply
\be
 h^{ }_\nu = - i 
 \sqrt{M}\, 
  R^{ }_{ }\, 
  P^{ }_{ }\,
  \sqrt{m^{ }_\nu}\, 
  U^{\dagger}_\rmii{PMNS}\, 
  \frac{\sqrt{2}}{v}
  \;, \la{soln_ci_2}
\ee
where the projection operator is 
\be
 P^{ }_\rmii{NH} = 
 \left( 
 \begin{array}{ccc}
   0 & 1 & 0 \\ 
   0 & 0 & 1 
 \end{array}
 \right)
 \;, \quad
 P^{ }_\rmii{IH} = 
 \left( 
 \begin{array}{ccc}
   1 & 0 & 0 \\ 
   0 & 1 & 0 
 \end{array}
 \right)
 \;. 
\ee
Only the phase
$\phi^{ }_1$ defined as in \eq\nr{phase}, present with both of the 
mass structures in \eqs\nr{mnu_nh} and \nr{mnu_ih}, 
is assumed non-zero.
In total there are 6 independent real parameters: 
$M^{ }_{1}$, $M^{ }_2$, 
$\re z$, $\im z$, $\delta$  and $\phi^{ }_1$. 
Benchmark values are  
given in \eqs\nr{params1}, \nr{params2}. 

%
\section{Parametrization of the Chern-Simons diffusion rate}

For the Chern-Simons diffusion rate we employ a numerical parametrization
based on classical lattice gauge theory simulations~\cite{sphaleron}. 
At low temperatures, the rate is approximated as 
\be
 \Gamma^{(T < \Tc^{ })}_\rmi{diff}  \; \simeq \; 
 T^4 \, \exp \biggl( -147.7 + \frac{0.83 T}{\mbox{GeV}}
 \biggr)
 \;. 
\ee
At high temperatures,  
$
 \Gamma^{(T > \Tc^{ })}_\rmi{diff} \; \simeq \;  
 18 \alphaw^5 T^4   
$. 
The rate originates from the diffusive Langevin dynamics of 
almost-static gauge fields~\cite{db}, and we therefore employ 
a dimensionally reduced gauge coupling for numerical estimates, 
\be 
 \alphaw^{ } \; \equiv \; \frac{g_\rmii{DR}^2}{4\pi}
 \;, \quad 
 g_\rmii{DR}^2 \; \approx \;  
 g_\rmi{w}^2 (\bmu) \, 
 \biggl\{ 
   1 + \frac{g_\rmi{w}^2(\bmu)}{(4\pi)^2}
  \biggl[ 
   \frac{43}{3} \ln \biggl( 
    \frac{\bmu e^{-\gammaE}}{4\pi T}
   \biggr)
  - 8 \ln \biggl( 
    \frac{\bmu e^{-\gammaE}}{\pi T}
   \biggr) + \frac{2}{3} 
  \biggr] 
 \biggr\} 
 \;, \la{alphaw}
\ee
where the $\msbar$ coupling is 
$
 g_\rmi{w}^2 (\bmu) \approx 48\pi^2 / 
 [19 \ln(\bmu / \Lambdamsbar) ]
$,
and we set $\bmu \simeq 2\pi T$ in practice.  
The value of $\Lambdamsbar$ is fixed by 
$g_\rmi{w}^2(\mZ) = 0.425$. 
The crossover from the high-temperature to the low-temperature
behaviour is rapid according to ref.~\cite{sphaleron}, 
and we have approximated it as  
\be 
 \Gamma^{ }_\rmi{diff}(T) \; \equiv \; \mathop{\mbox{min}}
 \bigl\{  
    \Gamma^{(T > \Tc^{ })}_\rmi{diff} , 
    \Gamma^{(T < \Tc^{ })}_\rmi{diff}
 \bigr\} 
 \;. 
\ee

\small{
%

}

\end{document}